\documentclass[fleqn,usenatbib]{mnras}

\usepackage{newtxtext,newtxmath}
\usepackage[T1]{fontenc}
\usepackage{amsmath}
\usepackage{booktabs}
\usepackage{xcolor}
\usepackage{hyperref}
\usepackage{natbib}
\usepackage{float}
\usepackage{placeins}

\DeclareRobustCommand{\VAN}[3]{#2}
\let\VANthebibliography\thebibliography
\def\thebibliography{\DeclareRobustCommand{\VAN}[3]{##3}\VANthebibliography}
\usepackage{comment}
\usepackage{titlesec}
\usepackage{caption}
\usepackage{graphicx}
\usepackage{tikz}
\usetikzlibrary{shapes.geometric, arrows.meta, positioning, fit, backgrounds}
\usepackage[normalem]{ulem}

\title[r-process and kilonova from magnetized AIC]{Collapse of Magnetized White Dwarfs as site of Heavy Element Formation and Kilonova Signal}

\author[Pitik et al.]{
	Tetyana Pitik$^{1}, $\thanks{E-mail: tetyana.pitik@berkeley.edu}
    David Radice$^{2,3,4}$,
    Daniel Kasen$^{1,5,6}$,
    Fabio Magistrelli$^{7}$,
    Patrick Chi-Kit Cheong$^{1}$, and
    \newauthor\ Sebastiano Bernuzzi$^{7}$
	\\
	$^{1}$ Department of Physics, University of California Berkeley, Berkeley, California 94720, USA\\
    $^{2}$ Institute for Gravitation and the Cosmos, The Pennsylvania State University, University Park, PA 16802, USA\\
    $^{3}$ Department of Physics, The Pennsylvania State University, University Park, PA 16802, USA\\
    $^{4}$ Department of Astronomy \& Astrophysics, The Pennsylvania State University, University Park, PA 16802, USA\\
    $^{5}$ Department of Astronomy, University of California Berkeley, Berkeley, California 94720, USA\\
    $^{6}$Nuclear Science Division, Lawrence Berkeley National Laboratory, Berkeley, CA 94720, USA\\
    $^{7}$ Theoretisch-Physikalisches Institut, Friedrich-Schiller-Universit\"at Jena, 07743, Jena, Germany
    }

\date{Accepted XXX. Received YYY; in original form ZZZ}

\pubyear{\the\year{}}

\begin{document}
\label{firstpage}
\pagerange{\pageref{firstpage}--\pageref{lastpage}}
\maketitle

\begin{abstract}
We present the first end-to-end calculation connecting the accretion-induced collapse (AIC) of a magnetized, rapidly rotating white dwarf to observable kilonova signatures, combining 2D general-relativistic neutrino-magnetohydrodynamic simulations, followed by radiation hydrodynamics with in-situ nuclear network and 2D Monte Carlo radiative transfer with spatially resolved heating rates.
Unlike all previous unmagnetized AIC models---which predicted proton-rich, $^{56}$Ni-dominated ejecta---strong magnetic fields eject ${\approx}\,0.2\,M_\odot$ of neutron-rich material ($\langle Y_e \rangle \sim 0.24$) on dynamical timescales, before neutrino irradiation can raise the electron fraction, enabling strong
$r$-process nucleosynthesis up to and beyond the third peak.
The resulting kilonova is lanthanide-rich ($X_{\rm lan} \approx 8\%$) and dominated by near-infrared emission.
We compute synthetic light curves in the LSST and JWST bands and find striking agreement, without parameter tuning, between the observations of AT~2023vfi/GRB~230307A and our broadband light curves for polar viewing angles.
These results establish magnetized AIC as a viable channel for heavy $r$-process element production and a compelling progenitor candidate for long-duration gamma-ray bursts with kilonova signatures.
\end{abstract}




\section{INTRODUCTION}

\begin{figure*}
	\centering
	\begin{tikzpicture}[
		stagebox/.style={
			rectangle,
			rounded corners=3pt,
			draw=black,
			line width=1pt,
			fill=blue!8,
			text width=4.2cm,
			minimum height=1.8cm,
			inner sep=6pt,
			align=center,
			font=\small,
			execute at begin node={\hyphenpenalty=10000}
		},
		stageboxgreen/.style={
			rectangle,
			rounded corners=3pt,
			draw=black,
			line width=1pt,
			fill=green!12,
			text width=4.2cm,
			minimum height=1.8cm,
			inner sep=6pt,
			align=center,
			font=\small,
			execute at begin node={\hyphenpenalty=10000}
		},
		stageboxpink/.style={
			rectangle,
			rounded corners=3pt,
			draw=black,
			line width=1pt,
			fill=pink!40,
			text width=4.2cm,
			minimum height=1.8cm,
			inner sep=6pt,
			align=center,
			font=\small,
			execute at begin node={\hyphenpenalty=10000}
		},
		titlebox/.style={
			rectangle,
			rounded corners=2pt,
			draw=none,
			fill=blue!25,
			text width=4.2cm,
			inner sep=6pt,
			align=center,
			font=\small\bfseries,
			execute at begin node={\hyphenpenalty=10000}
		},
		titleboxgreen/.style={
			rectangle,
			rounded corners=2pt,
			draw=none,
			fill=green!30,
			text width=4.2cm,
			inner sep=6pt,
			align=center,
			font=\small\bfseries,
			execute at begin node={\hyphenpenalty=10000}
		},
		titleboxpink/.style={
			rectangle,
			rounded corners=2pt,
			draw=none,
			fill=pink!60,
			text width=4.2cm,
			inner sep=6pt,
			align=center,
			font=\small\bfseries,
			execute at begin node={\hyphenpenalty=10000}
		},
		arrow/.style={
			->,
			>=Stealth,
			line width=1.5pt,
			color=black!70
		}
		]

		\node[stagebox] (grmhd) at (0,0) {2D GR$\nu$MHD  simulation with \textbf{\texttt{Gmunu}}};
		\node[titlebox, above=0pt of grmhd.north, anchor=south] (title1) {Collapse and explosion dynamics\\($0-1$~s)};

		\node[stageboxgreen] (knec) at (5.8,0) {2D Ray-by-ray radiation hydrodynamics with \textbf{\texttt{kNECnn}} (\textbf{\texttt{SNEC}}+ in-situ nuclear network \textbf{\texttt{SkyNet}})};
		\node[titleboxgreen, above=0pt of knec.north, anchor=south] (title2) {Hydrodynamics + nucleosynthesis\\($1\,\text{s}-30\,\text{days}$)};

		\node[stageboxpink] (sedona) at (11.6,0) {2D Monte Carlo radiative transfer with \textbf{\texttt{Sedona}} + nuclear heating from \textbf{\texttt{SkyNet}}};
		\node[titleboxpink, above=0pt of sedona.north, anchor=south] (title3) {Light curves and spectra\\($6\,\text{h}-20\,\text{days}$)};

		\draw[arrow] (grmhd.east) -- (knec.west);
		\draw[arrow] (knec.east) -- (sedona.west);
		
	\end{tikzpicture}
	\caption{Schematic overview of our computational pipeline for modeling AIC transient. The simulation proceeds through three stages: (1) GR$\nu$MHD evolution of the WD collapse and explosion with \texttt{Gmunu}, (2) ray-by-ray radiation hydrodynamics coupled with in-situ nucleosynthesis using \texttt{SNEC}+\texttt{SkyNet}, and (3) Monte Carlo radiative transfer with \texttt{Sedona} to produce synthetic observables.}
	\label{fig:pipeline}
\end{figure*}
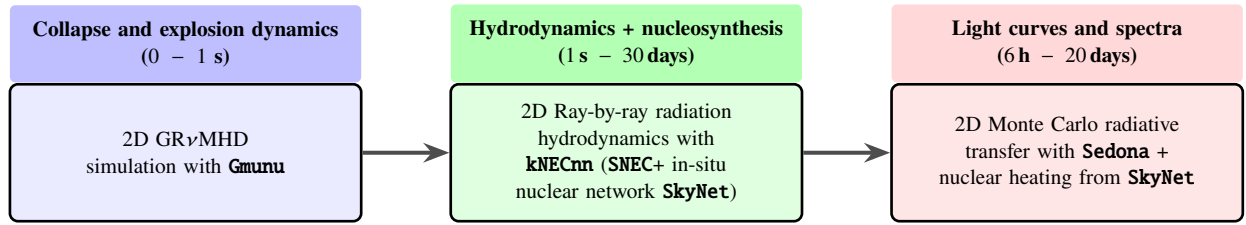

The accretion-induced collapse (AIC) of white dwarfs (WDs) represents a distinct pathway to neutron star (NS) formation, fundamentally different from the core collapse of massive stars~\citep{Ruiter:2018ouw,Wang:2020pzc}. There is some general consensus that when an oxygen-neon-magnesium (ONeMg) WD approaches the Chandrasekhar mass limit through accretion from a companion star, electron captures on $^{24}$Mg and $^{20}$Ne trigger gravitational collapse, resulting in the formation of a NS. The outcome of WD accretion depends critically on both the accretion rate and the initial WD mass~\citep{1986PrPNP..17..249N}, but the exact burning criterion that determines what actually occurs in nature is still a subject of debate~\citep{2005A&A...435..967Y,2016MNRAS.463.3461S,Kirsebom:2019tjd}. A similar outcome can occur through the merger of two WDs (merger-induced collapse, MIC), where the combined mass exceeds the stability threshold~\citep{1985A&A...150L..21S, 2006MNRAS.368L...1L}. For simplicity, we refer to both scenarios collectively as AIC throughout this work.

AICs are distinguished from ordinary core-collapse supernovae by several key properties. In case of unmagnetized WDs, the absence of an extended stellar envelope results in relatively low ejecta masses ($M_{\rm ej} \sim 10^{-3}$--$10^{-2}\,M_\odot$), modest explosion energies ($E_{\rm exp} \lesssim 10^{50}$~erg), and high ejecta velocities ($v \sim 0.1c$)~\citep{2006ApJ...644.1063D}. These characteristics make AICs appealing candidates for explaining rapidly evolving optical transients, including fast blue optical transients (FBOTs, see e.g.~\citealt{Govreen-Segal:2026skw}). Furthermore, if the collapsing WD is rapidly rotating and sufficiently magnetized, the resulting proto-neutron star (PNS) may form a magnetar capable of powering relativistic jets and gamma-ray bursts (GRBs; \citealt{2011MNRAS.413.2031M, Metzger:2018szx,Cheong:2024hrd}).

Recent observations have strengthened the connection between AICs and observable transients. The discovery of long-duration GRBs associated with kilonova-like emission, such as GRB~211211A~\citep{2022Natur.612..223R,2022Natur.612..232Y} and GRB~230307A~\citep{2024Natur.626..737L}, challenges the traditional association of long GRBs with massive star collapse and suggests alternative progenitors including compact object mergers and AICs. 

The theoretical understanding of AIC has developed substantially over the past decades. Early hydrodynamic simulations by~\citet{1987ApJ...320..304B} and~\citet{1992ApJ...391..228W} established that WD collapse produces weak explosions with modest ejecta masses. \citet{1999ApJ...516..892F} explored nucleosynthesis in AIC ejecta, finding conditions potentially favorable for $r$-process element production. \citet{2006ApJ...644.1063D} performed multi-dimensional AIC simulations with neutrino transport, and~\citet{2007ApJ...669..585D} showed that adding magnetic fields converts the weak neutrino-driven explosion into a stronger, magnetically driven one, increasing ejected mass and producing neutron-rich outflows favorable for $r$-process nucleosynthesis, while~\citet{Abdikamalov:2009aq} performed axisymmetric general relativistic simulations and predicted gravitational wave signals detectable from galactic AIC events. More recently,~\citet{LongoMicchi:2023khv} conducted 3D general relativistic simulations with neutrino transport, characterizing the multimessenger emission from non-magnetized AICs.

Recent work has highlighted the crucial roles of rotation and magnetic fields. In~\citet{Cheong:2024hrd}, we performed the first seconds-long 2D general-relativistic neutrino-magnetohydrodynamic simulations of AIC, revealing a qualitatively different picture from all previous studies: strongly magnetized, rapidly rotating WDs launch collimated relativistic jet and ejecta up to ${\sim}\,0.2\,M_\odot$ of neutron-rich material---one to two orders of magnitude more than in unmagnetized models---with electron fractions low enough for heavy $r$-process nucleosynthesis and explosion energies consistent with both short GRBs and kilonova-producing events.
This finding underscores the essential role of magnetic fields: two-dimensional neutrino-hydrodynamic simulations by~\citet{Batziou:2024ory}, performed without magnetic fields, demonstrated that rotation alone significantly influences the ejecta composition, with rapidly rotating models producing increasingly neutron-rich outflows at late times; however, the average electron fractions ($\langle Y_e \rangle \sim 0.43$--$0.50$) remain too high for heavy $r$-process nucleosynthesis. Three-dimensional simulations by~\citet{Kuroda:2025iyj} confirmed that rotation enhances both the ejecta mass and the neutron-richness of the material, while producing strong gravitational wave signals. Subsequent 3D GRMHD simulations by~\citet{Combi:2025yvs} showed that even starting from sub-surface seed fields, MRI-driven turbulent and mean-field $\alpha\Omega$ dynamo can provide sufficient large-scale flux to generate a magnetic tower capable of a successful jet-driven magnetorotational explosion. The convergence of these independent approaches---reaching similar outcomes of jet-driven explosions, neutron-rich outflows, and conditions consistent with GRBs and kilonovae---provides strong support for AIC as a viable progenitor of these transients.

Despite these advances, a critical gap remains in connecting the dynamical simulations to observable electromagnetic signatures. While previous studies have characterized the thermodynamic properties of AIC ejecta, the detailed nucleosynthetic yields and their imprint on kilonova light curves and spectra have not been computed self-consistently. Such calculations require explicit nuclear reaction network calculations that track the full $r$-process evolution rather than relying on parametric prescriptions, and frequency-dependent radiative transfer that captures the wavelength-dependent opacity contributions from the synthesized heavy elements.

In this work, we present a comprehensive end-to-end pipeline that connects GR$\nu$MHD simulations of AIC to observable electromagnetic signatures. Building upon 2D GR$\nu$MHD simulation performed with the \texttt{Gmunu} code~\citep{2021MNRAS.508.2279C,2023ApJS..267...38C}, we extract ejecta profiles and evolve them through the following stages: (1) radiation-hydrodynamic evolution with the \texttt{SNEC} code, coupled in-situ with the nuclear reaction network \texttt{SkyNet}~\citep{Lippuner:2017tyn,Magistrelli:2025xja} to compute nucleosynthetic yields; and (2) multi-dimensional Monte Carlo radiative transfer with the \texttt{Sedona} code~\citep{Kasen:2006ce,Roth:2014wda}, which we have extended to incorporate spatially and temporally resolved heating rates derived directly from the nucleosynthesis calculations. This approach represents a significant advance over previous studies that either assumed parametric nuclear heating or employed gray radiative transfer.

The structure of this paper is as follows. In Section~\ref{sec:methods}, we describe the numerical methods and the coupling between the various codes in our pipeline. Section~\ref{sec:results} presents our results, including the nucleosynthetic yields, heating rates, and synthetic light curves and spectra. In Section~\ref{sec:discussion}, we discuss our results in the context of previous AIC studies. We summarize our conclusions in Section~\ref{sec:conclusions}.

\section{METHODS}
\label{sec:methods}

Our computational pipeline consists of three stages: (1) general-relativistic magnetohydrodynamic (GR$\nu$MHD ) simulations of the AIC event using \texttt{Gmunu}, (2) radiation-hydrodynamic evolution with in-situ nuclear reaction network calculations using \texttt{SNEC} coupled to \texttt{SkyNet}, and (3) Monte Carlo radiative transfer using \texttt{Sedona}. In this section, we describe each component and the interfaces between them. A schematic overview of our pipeline is shown in Fig.~\ref{fig:pipeline}.

\subsection{GR$\nu$MHD simulation with \texttt{Gmunu}}
\label{sec:gmunu}

This work extends our previous investigation~(\citealt{Cheong:2024hrd}; CH2024 henceforth), which demonstrated that rapidly rotating, strongly magnetized WDs undergoing AIC produce neutron-rich outflows and potentially lead to a relativistic jet. 
Here, we briefly summarize the numerical setup and highlight the key modifications introduced in the present study.

We construct the initial WD model using the \texttt{RNS} code~\citep{1995ApJ...444..306S}, assuming rigid rotation and employing the finite-temperature LS220 equation of state~\citep{1991NuPhA.535..331L}. The progenitor has a gravitational mass of $1.5\,M_{\odot}$ and a central energy density of $\epsilon_{c}/c^2 = 10^{10}~\rm{g \cdot cm^{-3}}$, consistent with conditions expected to trigger electron-capture-induced collapse~\citep{1991ApJ...367L..19N, 2004A&A...419..623Y, 2006ApJ...644.1063D}. The initial thermodynamic state is set to $T = 0.01~\mathrm{MeV}$ and $Y_e = 0.5$. To quantify the rotational flattening, we use the axis ratio $a_r$, defined as the ratio of polar to equatorial radius. Motivated by the expectation that AIC progenitors spin rapidly as a consequence of prior accretion~\citep{2003ApJ...583..885P, 2004ApJ...615..444S, 2018ApJ...869..140K}, we adopt $a_r=0.75$, corresponding to $\Omega \approx 5~{\rm Hz}$ (roughly $80\%$ of the mass-shedding limit).

The WD is threaded by a magnetic field with poloidal and toroidal components, such that $B_{\rm{pol}}=B_{\rm{tor}}=10^{12}$~G, approximately uniform within a characteristic radius $r_{0}=600$~km, with poloidal component aligned with the spin axis~\citep{2007ApJ...669..585D,2021MNRAS.508.6033V}. This field strength is chosen to compensate for magnetic field amplification mechanisms---such as the magnetorotational instability (MRI) and turbulent dynamos~\citep{2006PhRvD..73j4015D, 2015Natur.528..376M}---that operate in the post-bounce environment but cannot be captured in our axisymmetric setup. This choice is further supported by the 3D GRMHD simulations of~\citet{Combi:2025yvs}, which showed that MRI-driven dynamos starting from buried seed field amplify the magnetic field to macroscopic strengths comparable to our initial configuration. Prior to evolution, we impose a temperature stratification $T = T_c \left( \rho / \rho_c \right)^{0.35}$ with $T_c = 5 \times 10^{9} ~{\rm K}$ ($\approx 0.43~{\rm MeV}$), following~\citet{2006ApJ...644.1063D,2007ApJ...669..585D}.

The dynamical evolution is performed with \texttt{Gmunu}~\citep{2021MNRAS.508.2279C,2023ApJS..267...38C}, a GR$\nu$MHD code that evolves the Einstein equations under the conformally flat approximation together with the equations of ideal MHD. Neutrino effects are incorporated via the energy-integrated two-moment transport scheme described in~\citet{2022MNRAS.512.1499R}, with interaction rates supplied by the \texttt{WeakRates} module of \texttt{WhiskyTHC}~\citep{2022MNRAS.512.1499R}. Magnetic flux conservation is enforced through staggered-mesh constrained transport~\citep{1988ApJ...332..659E}.

We perform axisymmetric simulations in cylindrical coordinates $(r, z)$. Unlike CH2024, where the domain spanned $2000$~km in each direction with north-south symmetry assumed, here we use a substantially larger grid covering $0 \leq r\leq 3 \times 10^{5}~{\rm km}$ and $-3 \times 10^{5} \leq z \leq 3 \times 10^{5}~{\rm km}$. This expanded domain guarantees that all unbound material remains within the computational volume at the extraction time ($t \sim 1$~s post-bounce). Crucially, we evolve both hemispheres independently without imposing equatorial symmetry, enabling us to capture potential asymmetries arising from the coupled evolution of rotation, magnetic stresses, and neutrino interaction. The computational grid consists of $128 \times 256$ base cells with a logarithmic radial grid and 14 levels of adaptive mesh refinement, yielding a minimum vertical spacing of $\Delta z \approx 286~{\rm m}$ near the stellar center when densities exceed
$10^{12}~{\rm g \cdot cm^{-3}}$. We adopt the HLL approximate Riemann solver~\citep{harten1983upstream}, third-order PPM spatial reconstruction~\citep{1984JCoPh..54..174C}, and the IMEXCB3a implicit-explicit time integrator~\citep{2015JCoPh.286..172C}. The LS220 equation of state~\citep{1991NuPhA.535..331L} is employed throughout the evolution.

Electron fraction evolution during infall follows the parametrized deleptonization prescription of~\citet{2005ApJ...633.1042L}. At core bounce, we activate the full two-moment neutrino transport and enable neutrino-matter coupling as detailed in~\citet{2024ApJ...975..116C}. We continue the evolution until ${\sim}1$~s after bounce, a characteristic timescale necessary to complete heavy $r$-process nucleosynthesis.

We note that self-consistent pre-collapse WD models from stellar evolution, in particular with different amounts of rotation, are not currently available; we therefore adopt a constructed initial model, following the standard approach of all previous AIC simulations~\citep{2006ApJ...644.1063D,2007ApJ...669..585D,Abdikamalov:2009aq,LongoMicchi:2023khv,Batziou:2024ory,Kuroda:2025iyj}, and encourage the community to provide such models, as they would
significantly improve the fidelity of AIC simulations. Neither the prior accretion phase nor the deflagration that may precede the collapse are modelled: whether the WD collapses or undergoes thermonuclear ignition depends on the central density, accretion rate, and flame physics~\citep{2005A&A...435..967Y,2016MNRAS.463.3461S,Kirsebom:2019tjd,Wang:2020pzc}. Recent 3D simulations by~\citet{2026A&A...707A..84H} find that the transition between thermonuclear explosion and gravitational collapse occurs in the range $\log_{10}\rho_c \approx 10.0$--$10.15~\mathrm{g\,cm^{-3}}$, with collapse favoured at the higher densities; our initial central density of $10^{10}~\mathrm{g\,cm^{-3}}$  falls within this transition regime, consistent with collapse being a plausible outcome. Our assumed $Y_e = 0.5$ is consistent with the mass-weighted average $Y_e = 0.493$ obtained from \texttt{MESA} stellar evolution calculations of ONe WD progenitors~\citep{2013ApJ...772..150J,2016A&A...593A..72J}. In any case, the collapsing object should be essentially independent of its evolutionary history, the only factors that matter being the final mass, angular momentum, and central density. Although the LS220 EOS assumes NSE throughout, in the cold, degenerate WD core the pressure is dominated by relativistic degenerate electrons and is insensitive to the detailed nuclear composition; once the temperature rises during collapse and NSE genuinely holds, the composition is determined entirely by the local $(\rho, T, Y_e)$ and all memory of the pre-collapse nuclear species is erased. Composition-resolved EOSs are important for modelling the pre-collapse deflagration physics, but since we begin directly from the collapse, a nuclear-matter EOS is appropriate for our simulation. The parametrized deleptonization $\bar{Y}_e(\rho)$ trajectory used during infall was computed with the same LS220 EOS, ensuring self-consistency between the EOS and the electron-capture treatment, which is only active until core bounce.

\subsection{Hydrodynamics and Nucleosynthesis with \texttt{kNECnn}}
\label{sec:nucleosynthesis}

To interface the GR$\nu$MHD output with the subsequent nucleosynthesis calculations, we extract radial profiles of thermodynamic quantities along discrete angular directions. We sample 35 polar angles from $\theta = 5^\circ$ to $\theta = 175^\circ$ in increments of $\Delta\theta = 5^\circ$, where $\theta$ is measured from the positive $z$-axis. For each angle, we interpolate the rest-mass density $\rho$, temperature $T$, radial velocity $v_r$, electron fraction $Y_e$, and specific entropy $s$ as functions of the enclosed ejecta mass $m(r) = \int_{r_{\rm{cut}}}^r 4\pi r'^2 \rho(r') \, dr'$, where $r_{\rm{cut}}\sim 400$~km.

Following the extraction, each angular section is treated as a 1D spherical problem and evolved through the nucleosynthesis epoch using \texttt{kNECnn}~\citep{Magistrelli:2024zmk, Magistrelli:2025xja}, an updated version of the KiloNova Explosion Code \texttt{KNEC}~\citep{Wu:2021ibi}, which is itself based on the SuperNova Explosion Code \texttt{SNEC}~\citep{Morozova:2015bla}. \texttt{SNEC} is a spherically symmetric 1D Lagrangian radiation-hydrodynamics code that solves the equations of mass, momentum, and energy conservation, with radiation transport treated in the flux-limited diffusion approximation. The Lagrangian formulation, in which the enclosed mass $m$ serves as the independent spatial coordinate, is well-suited for coupling with nuclear reaction networks that require following the thermodynamic history of each mass element.

Whereas the original \texttt{SNEC} assumes $^{56}$Ni-powered heating appropriate for supernovae, \texttt{kNECnn} replaces this with a direct, in-situ coupling to the nuclear reaction network \texttt{SkyNet}~\citep{Lippuner:2017tyn}. \texttt{SkyNet} includes 7836 isotopes ranging from free nucleons to $^{337}$Cn, connected by ${\sim}140{,}000$ reactions. The full set of nuclear-physics inputs (reaction rates, masses, partition functions, weak rates, fission and $\beta$-delayed neutron emission prescriptions) is identical to that of~\citet{Lippuner:2017tyn}, to which we refer the reader for details. Each Lagrangian mass shell independently evolves its own instance of the nuclear network, synchronized with the hydrodynamic timestep. At each hydrodynamic time step, the network, which has its own sub-steps based on the nuclear timescales, receives the local thermodynamic conditions $(\rho, T)$ and returns the updated composition and nuclear energy generation rate, $\dot{q}_{\rm nucl}$. This enters the Lagrangian energy equation:
\begin{equation}
    \frac{\partial \epsilon}{\partial t} = \frac{P}{\rho}\frac{\partial \ln \rho}{\partial t} - 4\pi r^2 Q \frac{\partial v}{\partial m} - \frac{\partial L}{\partial m} + \dot{q}_{\rm nucl},
    \label{eq:energy}
\end{equation}
where $\epsilon$ is the specific internal energy, $P$ and $\rho$ are the pressure and density, $r$ and $v$ are the radial position and velocity, $m$ is the Lagrangian mass coordinate, $Q$ is the artificial viscosity, $L$ is the radiative luminosity, and $\dot{q}_{\rm nucl}$ is the thermalized nuclear heating rate. See~\citet{Magistrelli:2024zmk, Magistrelli:2025xja} for more details on the coupling infrastructure.

We note that \texttt{kNECnn} does not include neutrino absorption reactions: the electron fraction handed off from \texttt{Gmunu} serves as the initial condition and is subsequently modified only by the weak decays internal to the network (e.g.\ $\beta^{-}$ decays along the $r$-process path). Although \texttt{SkyNet} supports neutrino interactions on free nucleons~\citep{Lippuner:2017tyn}, these require external neutrino luminosities and spectra as input, which are not provided in our setup. This omission is justified, since the extracted ejecta profiles are well beyond the neutrinosphere ($R_\nu \lesssim 80$~km), where the neutrino flux drops by orders of magnitude, and more fundamentally, the magnetically-driven ejection operates on a dynamical timescale that is shorter than the neutrino re-leptonization timescale: material is launched before neutrino irradiation can raise $Y_e$ to weak equilibrium, which is the central mechanism that enables heavy $r$-process nucleosynthesis in our model.

At the extraction time ($t \sim 1$~s) of the profile from \texttt{Gmunu}, many mass shells, especially in the low density tail, have already cooled to temperatures too low for nuclear statistical equilibrium (NSE) to hold. Since tracers are not modeled in our simulation, we rely on the NSE assumption to estimate the initial isotopic composition, and use the backtracking-NSE initialization at $T_{\rm NSE}=8$~GK, as described in~\citep{Radice:2016dwd,Magistrelli:2025xja}. The backtracking procedure is purely an initialization step and does not generate any nuclear energy. Shells that are still above $T_{\rm NSE}$ at the extraction time are in NSE and are initialised directly from their current $(\rho, T, Y_e,s)$. For shells that have already cooled below $T_{\rm NSE}$ at extraction, the thermodynamic trajectory is extrapolated backward (assuming homologous expansion) to find the last time the shell was at $T_{\rm NSE}$, and NSE is imposed there; the network is then advanced forward on the actual thermodynamic trajectory to capture the full transition from NSE to the $r$-process regime. The systematic uncertainties associated with this initialization scheme are discussed in Appendix~\ref{app:initialization}.

The radioactive decay of $r$-process nuclei releases energy through $\gamma$-rays, $\beta$-decay electrons and positrons, $\alpha$ particles, fission fragments, and neutrinos. Only a fraction of this energy thermalizes locally; the remainder escapes.  The thermalized heating rate that enters the energy equation~\ref{eq:energy} and drives the kilonova emission can be thus expressed as
\begin{equation}
    \dot{q}_{\rm nucl} = \eta_\gamma \dot{q}_\gamma + \eta_{e^\pm} \dot{q}_{e^\pm} + \eta_\alpha \dot{q}_\alpha + \eta_{\rm fiss} \dot{q}_{\rm fiss},
\end{equation}
where $\eta_{i}$ represent the thermalization efficiency of particle $i$.
The number and spectra of emitted $\gamma$-rays is computed combining the instantaneous composition of the ejecta with the lifetime and spectra from the \href{https://www-nds.iaea.org/exfor/endf.htm}{ENDF/B-VIII.0} database~\citep{2018NDS...148....1B}. The \href{https://www.nist.gov/pml/xcom-photon-cross-sections-database}{NIST XCOM} cross-sections are then used to calculate the $\gamma$ thermalization factor as described in \cite{Hotokezaka:2019uwo, Magistrelli:2025xja}.
Charged particles ($e^\pm$, $\alpha$) follow the analytic prescription of~\citet{Kasen:2018drm}; the energy carried away by neutrinos from $\beta$-decay reactions is subtracted from the total nuclear energy release inside the network, so that $\dot{q}_{\rm nucl}$ does not include the neutrino component---neutrino cooling is therefore fully accounted for; while for the residual nuclear power, arising from the recoil energy of daughter nuclei, fission fragments, (soft) X-rays, proton and neutron emission, and delayed electron (e.g., Auger) processes, we adopt a constant and spatially uniform thermalization factor $\eta_{\rm fiss} = 0.8$~\citep{Magistrelli:2025xja}. 

In our 2D ray-by-ray approach, each of the 35 angular profiles is evolved independently. We neglect non-radial flows between angular sections, an approximation justified for the predominantly radial expansion at late times. We also ignore radial composition mixing, which has been shown to have minimal effect on the total integrated abundances~\citep{Zhai:2026uwn}.
For each ray, we simulate the ejecta evolution for $\sim 30$~days, recording the density, temperature, and velocity profiles, isotopic abundances $X_i(m, \theta)$, and the time-dependent thermalized heating rates $\dot{q}_{\rm nucl}(t, m, \theta)$.

\subsection{Monte Carlo Radiative Transfer with \texttt{Sedona}}
\label{sec:sedona}

\begin{figure*}
	\centering
	\hspace{-8mm}
	\includegraphics[width=0.95\textwidth]{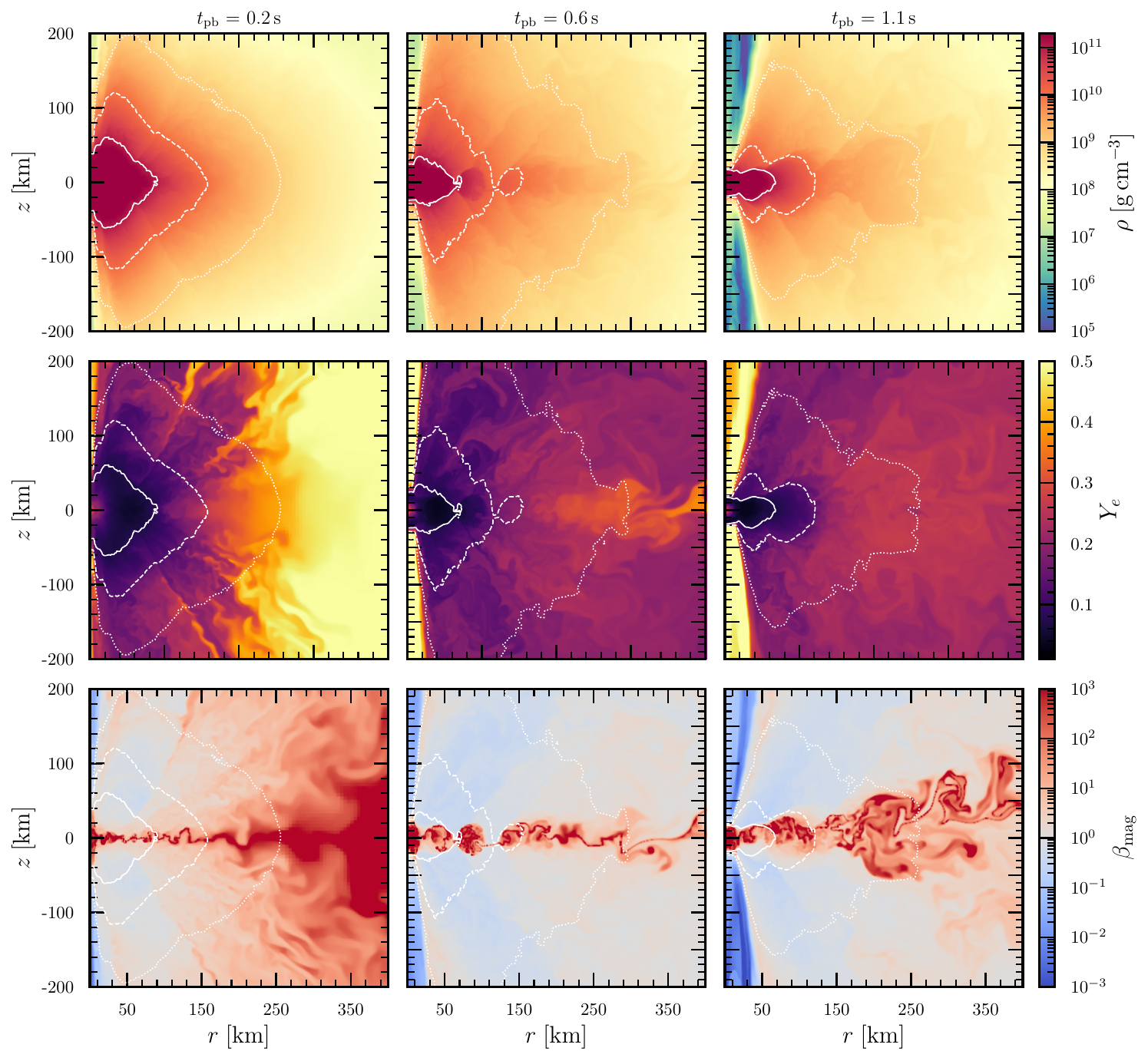}
	\caption{2D snapshots in the $rz$-plane at $t_{\rm pb} = 0.2$, $0.6$, and $1.1~{\rm s}$ (left to right), showing the rest-mass density $\rho$ (top), electron fraction $Y_e$ (middle), and plasma $\beta_{\rm mag} \equiv P_{\rm gas}/P_{\rm mag}$ (bottom). A magnetically-dominated ($\beta_{\rm mag} < 1$), low-density outflow develops along the polar axis, while the bulk of the neutron-rich material ($Y_e \lesssim 0.25$) is concentrated at mid-latitudes, between the polar outflow and the equatorial disk. Solid, dashed, and dotted lines represent density contours of $10^{11} \mathrm{g\, cm^{-3}},10^{10}\mathrm{g\, cm^{-3}}$ and $10^{9}\mathrm{g\, cm^{-3}}$, respectively. }
	\label{Fig: rho-ye-beta_mag}
\end{figure*}

Once the ejecta reach homologous expansion, we pass the output from \texttt{kNECnn} to the Monte Carlo radiative transfer code \texttt{Sedona}~\citep{Kasen:2006ce, Roth:2014wda} to compute synthetic kilonova light curves and spectra. \texttt{Sedona} models radiation as discrete packets of photon energy that propagate through the expanding medium, undergoing absorption, emission, and scattering events. Photons reaching the outer boundary escape and are binned by time, frequency, and viewing angle to generate angle-resolved spectral time series.

We extract the thermodynamic profiles (density $\rho$, temperature $T$, velocity $v$) from \texttt{kNECnn} at $t = 6$~hours post-bounce, when the ejecta are already in homologous expansion. For the composition, we extract the isotopic mass fractions $X_i(m, \theta)$ at $t = 10$~days. Since \texttt{Sedona} does not evolve the nuclear composition, we choose a late-time snapshot when the short-lived isotopes have decayed and the abundances are dominated by longer-lived species that persist through the epochs of interest ($t \lesssim 20-30$~days).

The 35 angular profiles from the ray-by-ray \texttt{kNECnn} simulations are combined into a 2D axisymmetric model on a cylindrical $(r, z)$ grid. We employ a cylindrical grid consisting of $80 \times 160$ cells in the $r$ and $z$ directions. 
For each grid cell, we interpolate the thermodynamic quantities from the 1D profiles based on the cell's polar angle ${\theta = \arctan(r/z)}$. Density, temperature, and composition are interpolated using mass-weighted averages across contributing angular profiles. 

\begin{figure*}
	\centering
	\hspace{-8mm}
	\includegraphics[width=0.85\textwidth]{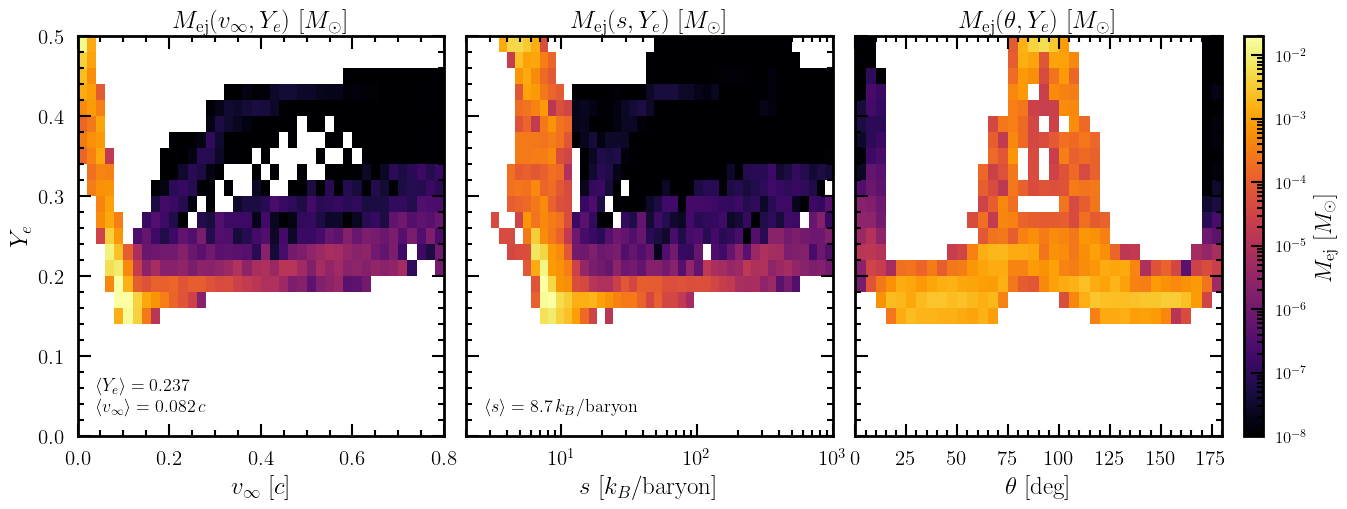}
	\caption{Joint distributions of ejecta mass at $t_{\rm{pb}} = 1.1$~s as a function of electron fraction $Y_e$ versus asymptotic velocity $v_\infty$ (left), specific entropy $s$ (center), and polar angle $\theta$ (right). The color scale encodes the ejected mass in each bin, revealing correlations between composition and kinematics: the bulk of the neutron-rich ($Y_e \lesssim 0.25$) material is concentrated at moderate velocities ($v_\infty \sim 0.05$--$0.3\,c$), low entropies ($s \lesssim 20\,k_B$/baryon), and polar latitudes. Mass-weighted averages are indicated in each panel.}
	\label{Fig: 2D hist}
\end{figure*}

\begin{figure*}
	\centering
	\hspace{-8mm}
	\includegraphics[width=1.02\textwidth]{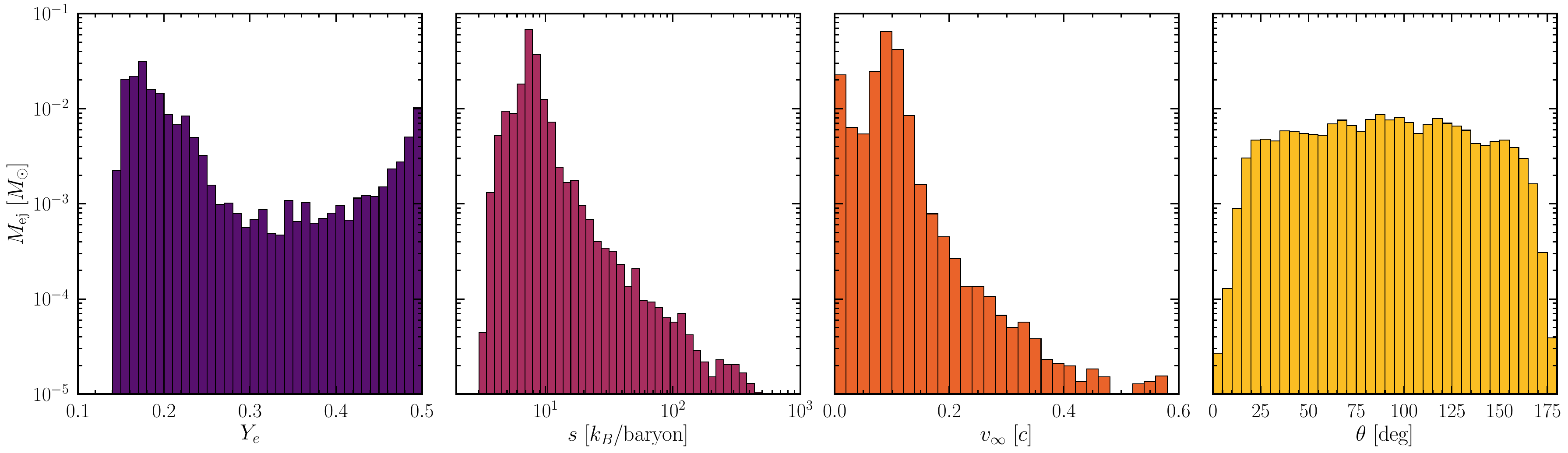}
	\caption{Mass distributions of the ejecta at $t_{\rm{pb}} = 1.1$~s as a function of electron fraction $Y_e$, specific entropy $s$, asymptotic velocity $v_\infty$, and polar angle $\theta$ (left to right). The distributions are obtained by integrating the 2D histograms of Fig.~\ref{Fig: 2D hist} over the complementary variable. The ejecta are dominated by neutron-rich material ($Y_e \sim 0.15$--$0.25$), with a secondary proton-rich component near $Y_e \sim 0.5$. Most of the mass has low entropy ($s \sim 3$--$20\,k_B$/baryon) and moderate velocities ($v_\infty \sim 0.05$--$0.2\,c$).}
	\label{Fig: 1D his}
\end{figure*}

A key contribution of this work is the implementation of spatially varying, time-dependent heating rates in \texttt{Sedona}. The standard \texttt{Sedona} heating prescriptions apply the same functional form to all grid cells. This approximation breaks down for AIC ejecta, where significant variations in properties lead to different nucleosynthesis products and, consequently, distinct heating histories.

We modified the \texttt{Sedona} source code to read per-cell heating rates from external tables generated by \texttt{kNEcnn}. 
For each grid cell $i$, we store the thermalized heating rate $\dot{q}_{\rm nucl}(t)$ as a function of time, tabulated up to $30$~days. At runtime, \texttt{Sedona} interpolates these tables in log-log space to obtain the heating rate at any given time.

Crucially, the heating rates passed to \texttt{Sedona} are already thermalized---\texttt{kNECnn} has computed the thermalization efficiencies for each particle type, as described in Sec.~\ref{sec:nucleosynthesis}.  Thus, \texttt{Sedona} receives the net heating that deposits into the ejecta thermal energy, without needing to recompute thermalization. This ensures consistency between the nucleosynthesis and radiative transfer calculations.

Matter ionization and excitation state are modeled under the assumption of local thermodynamic equilibrium (LTE),
which provides a good approximation while the ejecta remain optically thick but becomes less accurate at late times as the material becomes optically thin. We use theoretical atomic data from~\citet{Tanaka:2019iqp}, computed using the \texttt{HULLAC} code~\citep{2001JQSRT..71..169B}, merged with \href{https://sites.pitt.edu/~hillier/web/CMFGEN.htm}{\texttt{CMFGEN}} compilation of atomic data to provide energy levels and transition strengths for elements spanning $1 \leq Z \leq 87$. 
\texttt{HULLAC} models electron orbital functions using a single-electron Dirac equation with a central-field potential, constructed from spherically averaged electron–electron interactions and the nuclear charge. Because $p$, $d$, and $f$ orbitals are inherently aspherical, this approximation can lead to shifts in their energy levels.

Bound-bound opacities are treated in the Sobolev approximation~\citep{1960mes..book.....S} within an expansion opacity formalism, supplemented by free-free and electron-scattering contributions. \texttt{Sedona} calculates the opacity based on the detailed composition in each zone, capturing the wavelength-dependent absorption associated with each electronic transition. 

The 2D simulations produce angle-resolved spectra by binning escaping photon packets according to their direction of escape. With $N_\mu = 13$ angular bins in $\cos\theta$, we compute light curves and spectra at viewing angles from $\theta_{\rm obs} = 0^\circ$ (pole-on) to $180^\circ$ (opposite pole), capturing the variation that results from the latitude-dependent composition of the AIC ejecta.

\section{RESULTS}
\label{sec:results}

\subsection{Collapse and explosion dynamics}
\label{sec:gmunu results}

\begin{figure}
	\centering
	\includegraphics[width=1\columnwidth]{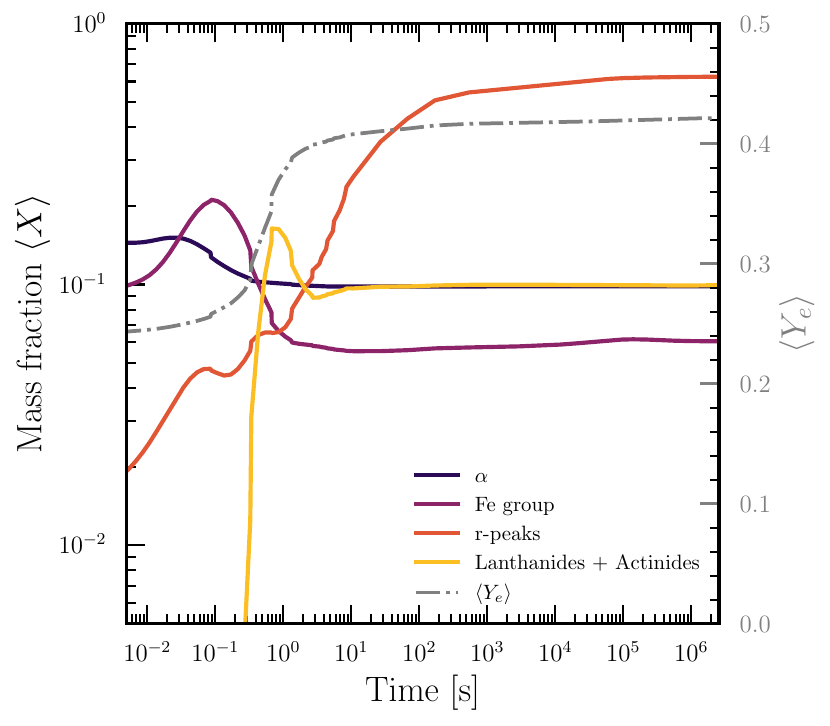}
	\caption{Time evolution of the mass-weighted average mass fractions $\langle X \rangle$ for grouped species: $\alpha$ particles, iron-group elements, $r$-process peak elements, and lanthanides + actinides. The right axis shows the mass-weighted average electron fraction $\langle Y_e \rangle$ (grey dash-dotted).}
	\label{Fig: mass_fraction_selected}
\end{figure}

The collapse dynamics, core bounce, and post-bounce evolution of the $B_{\rm pol} = B_{\rm tor}=10^{12}~{\rm G}$ model are fully consistent with those reported for the model $B_{\rm pol} = 10^{12}~{\rm G}$ in CH2024, to which we refer the reader for a detailed discussion.
Here we briefly summarize the key results relevant to the nucleosynthesis analysis that follows, and highlight the differences arising from the extended simulation domain and the relaxation of north--south symmetry.

Fig.~\ref{Fig: rho-ye-beta_mag} shows 2D snapshots of the rest-mass density, electron fraction, and  plasma $\beta_{\rm mag} \equiv P_{\rm gas} / P_{\rm mag}$ at $t_{\rm{pb}}\equiv t - t_{\rm bounce}= 0.2$, $0.6$, and $1.1$~s post-bounce.
Following core bounce, magnetic fields in the polar region are rapidly amplified by flux conservation and by magnetic winding.
The magnetization in the polar funnel drops well below unity already at early times ($t_{\rm{pb}} \lesssim 0.2~{\rm s}$) and continues to decrease, so that by the end of the simulation the magnetic pressure dominates the gas pressure throughout the jet region.
This progressively stronger magnetization drives an increasingly powerful collimated outflow along the polar axis, while simultaneously accelerating uncollimated ejecta at mid-latitudes.
Because the magnetically-driven ejection is fast enough to launch material before neutrino interactions can raise $Y_e$ to weak equilibrium, the outflow becomes increasingly neutron-rich over time: the $Y_e$ snapshots in Fig.~\ref{Fig: rho-ye-beta_mag} show that the neutron-rich component  grows in both mass and spatial extent as the simulation progresses. 

The overall morphology---a magnetically-dominated polar jet surrounded by neutron-rich equatorial outflows---is qualitatively unchanged with respect to CH2024.
The north and south hemispheres evolve nearly symmetrically, confirming that the imposed equatorial symmetry in CH2024 did not artificially bias the results.
The larger domain ($3 \times 10^5~{\rm km}$) allows us to follow the ejecta to greater distances, providing more reliable properties for the profiles handed off to \texttt{kNECnn}.

By the end of the simulation, at $t_{\rm{pb}} \lesssim 1.1~{\rm s}$, the total ejected mass is $M_{\rm ej} \approx 0.18~M_{\odot}$ with a kinetic energy of $E_{\rm ej} \simeq 8\times 10^{50}~{\rm erg}$, broadly consistent with the values reported in CH2024. The modest differences arise primarily from the ejecta extraction method: whereas CH2024 computed the ejecta as an integrated flux through a spherical surface at $1800~{\rm km}$, here we integrate over the entire computational domain at the final time step, which excludes material that has already fallen below the atmosphere floor. This leads to a slight underestimate of both mass and energy. 

The ejecta properties most relevant to nucleosynthesis---electron fraction, asymptotic velocity~\footnote{We estimate the asymptotic velocity as $v_{\infty} = \sqrt{1 - 1/\Gamma_{\infty}^2}$, where $\Gamma_{\infty} = - u_t h_{\rm tot}$, with $h_{\rm tot} = 1 + \varepsilon + P/\rho + b^2/\rho$ being the specific enthalpy with magnetic field contribution, $\varepsilon$ the internal energy, $P$ the pressure, $\rho$ the density, $b^2$ the contraction of the magnetic fields in the fluid frame, $h_{\min}$ is the minimum allowed enthalpy values for a given equation-of-state, $u_t = W(- \alpha + \beta_i v^i )$, where $v_r$ is the radial velocity, $W$ the Lorentz factos, $\alpha$ is the lapse function, and $\beta^{i}$ the space-like shift vector.}, and specific entropy---are illustrated in Figs.~\ref{Fig: 2D hist} and~\ref{Fig: 1D his}.
The angular distribution (Fig.~\ref{Fig: 2D hist}) reveals that the bulk of the neutron-rich material ($Y_e \lesssim 0.25$) is concentrated in the mid polar region, with typical velocities $v_{\infty} \lesssim 0.2\,c$ and entropies $s \lesssim 20\,k_B/{\rm baryon}$.
The fastest ejecta, found at the head of the outflow, consist primarily of high-$Y_e$ material.
The mass-weighted average properties are $\langle Y_e \rangle \sim 0.24$, $\langle v_{\infty} \rangle \sim 0.1\,c$, and $\langle s \rangle \sim 9\,k_B/{\rm baryon}$.
Fig.~\ref{Fig: 1D his} shows the mass distribution as a function of these quantities, integrated over the computational domain at the final time step.
In total, ${\sim}\,0.14~M_{\odot}$ of material has $Y_e \leq 0.25$, providing favorable conditions for $r$-process nucleosynthesis, which we explore in detail in Section~\ref{sec:skynet results}.

\begin{figure*}
	\centering
	\hspace{-8mm}
	\includegraphics[width=0.7\textwidth]{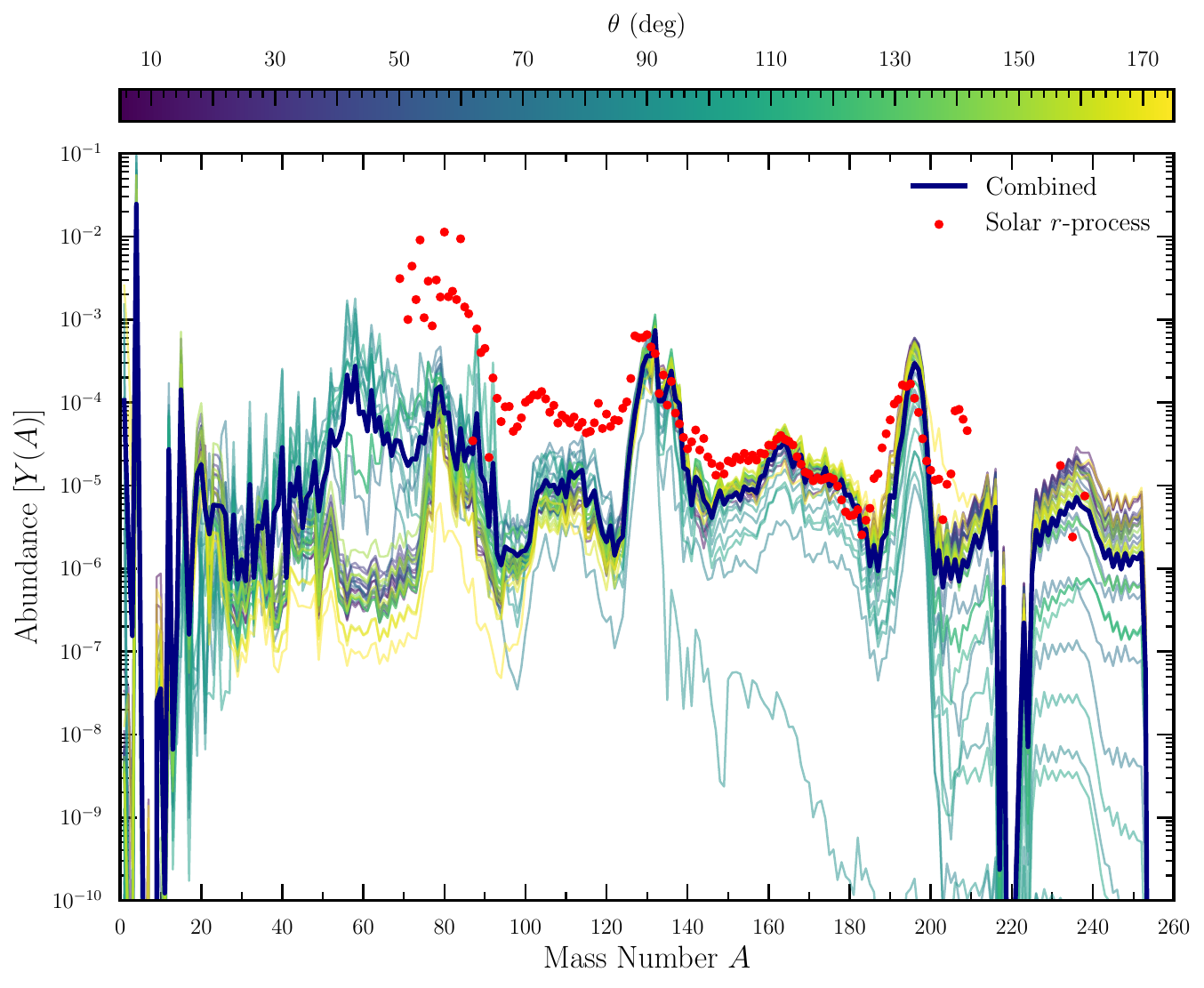}
	\caption{Mass-weighted average abundance pattern of the AIC ejecta (solid line) compared with the solar $r$-process residuals (circles;~\citealt{Prantzos:2019bpv}). Abundances are normalized to the second-peak region ($130 \leq A \leq 140$).
	}
	\label{Fig: abundaces}
\end{figure*}

\begin{figure*}
	\centering
	\hspace{-8mm}
	\includegraphics[width=1\textwidth]{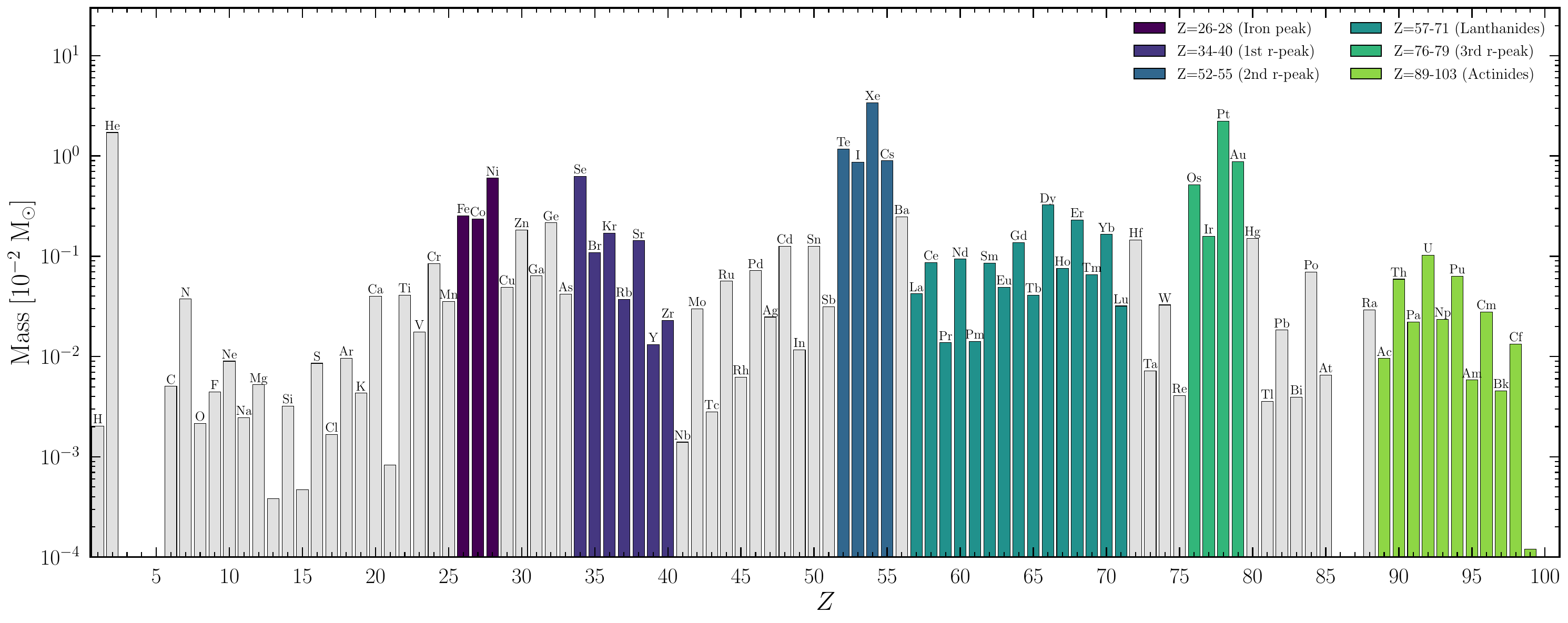}
	\caption{Total ejecta mass in each element as a function of atomic number $Z$, integrated over the ejecta at $t \simeq 30$~days. The four most abundant elements---Xe ($Z=54$), He ($Z=2$), Pt ($Z=78$), and Te ($Z=52$)---together constitute more than $50\%$ of the total ejecta mass.}
	\label{Fig: mass_Z}
\end{figure*}


\subsection{Nucleosynthesis}
\label{sec:skynet results}

The 35 angular profiles are evolved with \texttt{kNECnn} for ${\sim}\,30$~days, during which the in-situ nuclear network tracks the full nucleosynthesis and subsequent radioactive decays.
Because the ejecta span a wide range of electron fractions ($Y_e \sim 0.1$--$0.5$), the nucleosynthesis proceeds through qualitatively different channels depending on latitude: the neutron-rich mid-latitude material ($Y_e \lesssim 0.25$) undergoes  strong $r$-process nucleosynthesis, producing heavy elements up to and beyond the third $r$-process peak ($A \gtrsim 195$), while the higher-$Y_e$ material in the equatorial region synthesizes lighter species, including iron-group and first-peak elements.

Fig.~\ref{Fig: abundaces} compares the mass-weighted average abundance pattern of our ejecta with the solar $r$-process residuals~\citep{Prantzos:2019bpv}.
The abundances are normalized by fixing the cumulative fraction of elements in the range $170 \leq A \leq 200$ to match the solar values.
The agreement is remarkable across the second ($A \sim 130$) and third ($A \sim 195$) $r$-process peaks, as well as in the rare-earth region ($A \sim 160$) and the actinides ($A \gtrsim 230$), indicating that the AIC ejecta from a strongly magnetized progenitor can reproduce the heavy-element solar abundance pattern with high fidelity, while they strongly underproduce the first peak ($A \sim 80$) elements.
We note that the abundances shown here are extracted at $t \simeq 30$~days; the abundance pattern will continue to be shaped by the decay of longer-lived isotopes beyond this time.

Fig.~\ref{Fig: mass_Z} shows the total mass in each element as a function of atomic number $Z$, integrated over all angular profiles at the end of the simulation ($t \simeq 30$~days).
The four most abundant elements by mass are Xe ($Z = 54$), He ($Z = 2$), Pt ($Z = 78$), and Te ($Z = 52$), which together account for more than $50\%$ of the total ejecta mass.
The prominence of He reflects the $\alpha$-rich freeze-out from NSE in the high-$Y_e$ ejecta component, where $\alpha$ particles are not fully assembled into heavier nuclei. Fig.~\ref{Fig: mass_fraction_selected} shows the time evolution of grouped mass fractions: the $\alpha$ and iron-group components are set during the NSE freeze-out within $\sim 1$~s, lanthanides/actinides are assembled within the first $\sim 1$~s, while the $r$-process peak elements are built on longer timescales. The average $Y_e$ rises from $\sim 0.24$ to $\sim 0.42$ over $\sim 10$~s due to $\beta$-decays along the $r$-process path. The concentration of mass around Xe and Te (second $r$-process peak) and Pt (third peak) is characteristic of a robust $r$-process with a broad range of neutron-rich conditions.
The lanthanide mass fraction is $X_{\rm lan} \approx 8\%$ at $30$~days, and remains roughly constant during the relevant period of interest up to $t\sim 25$~days, indicating that the emitted light will be strongly reprocessed by their high bound-bound opacity, as we show in the next section.

\subsection{Kilonova signal}
\label{sec:sedona results}

\begin{figure*}
	\centering
	\hspace{-8mm}
	\includegraphics[width=0.9\textwidth]{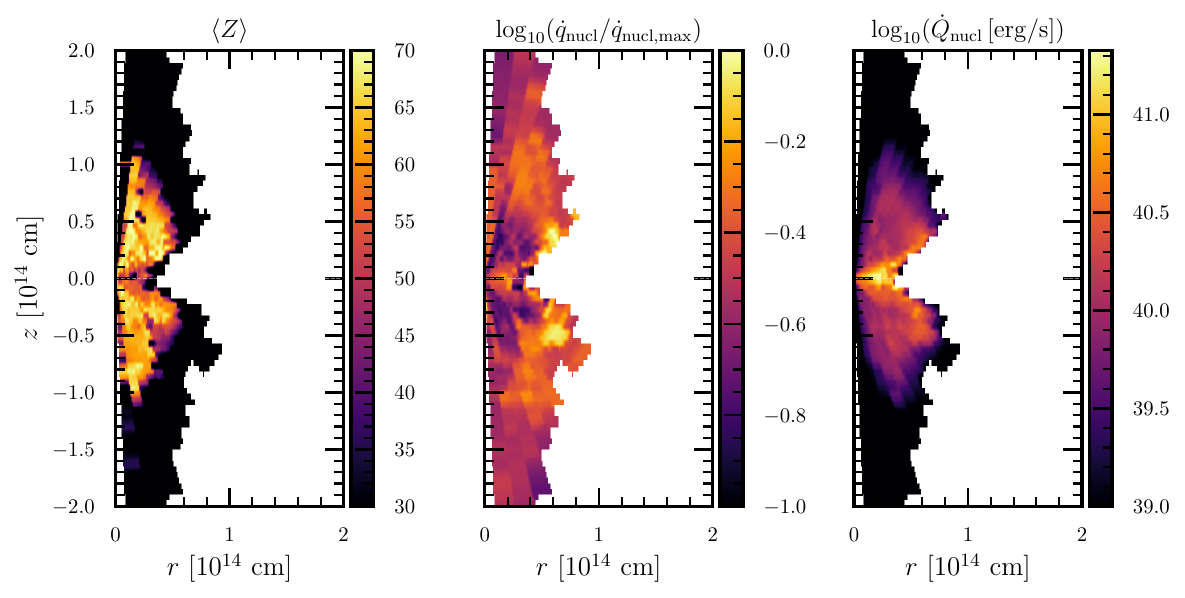}
	\caption{Ejecta structure in the $r$--$z$ plane at the \texttt{Sedona} extraction time ($t = 6$~h). \textit{Left:} mass-weighted average atomic number $\langle Z \rangle$, evaluated at $t = 10$~days after $r$-process freeze-out. The heaviest elements are concentrated in the inner, slower-moving core, while the fast outer ejecta are dominated by lighter species. \textit{Middle:} specific thermalized heating rate normalized to the maximum value at this time, $\dot{q}_{\rm nucl}/\dot{q}_{\rm nucl,max}$. The heating rate varies by up to an order of magnitude across the ejecta, underscoring the importance of spatially resolved heating in radiative transfer calculations. \textit{Right:} total nuclear luminosity per cell $\dot{Q}_{\rm nucl}$. Despite higher specific heating rates in the polar regions, the total luminosity is dominated by the equatorial ejecta, where the bulk of the mass resides.}
	\label{Fig: input_sedona_profile}
\end{figure*}

Fig.~\ref{Fig: input_sedona_profile} illustrates the ejecta structure at the \texttt{Sedona} extraction time ($t = 6$~h) in the $r$--$z$ plane, showing three quantities: the mass-weighted average atomic number $\langle Z \rangle$ (evaluated at $t = 10$~days, after $r$-process freeze-out; left panel), the specific thermalized heating rate normalized to its maximum value at that time ($\dot{q}_{\rm nucl}/\dot{q}_{\rm nucl,max}$; middle panel), and the total nuclear luminosity per cell ($\dot{Q}_{\rm nucl}$; right panel).

The left panel reveals a clear compositional stratification: the heaviest elements ($\langle Z \rangle \gtrsim 50$) are concentrated in the inner, slower-moving core, while the outermost, fastest ejecta are dominated by lighter species ($\langle Z \rangle \lesssim 35$).
This is fully consistent with the ejecta properties extracted from \texttt{Gmunu} (cf.\ the velocity--angle distribution in Fig.~\ref{Fig: 2D hist}), where the fastest material corresponds to the high-$Y_e$ material.

This stratification has direct consequences for the observed spectra: at early times, the photosphere recedes through the lighter outer layers, producing bluer emission, while at later epochs the lanthanide- and actinide-rich core increasingly dominates the opacity, driving the emission to the near-infrared.

The middle panel shows that the specific heating rate varies by up to an order of magnitude across the ejecta, reflecting the strong spatial dependence of the nucleosynthesis.
This substantial spatial variation confirms that a uniform heating prescription would introduce significant systematic errors in the predicted light curves and spectra.

The right panel shows the total nuclear luminosity per cell, which peaks in the equatorial region where the bulk of the ejecta mass is concentrated.
Although the polar ejecta have higher specific heating rates, the equatorial material dominates the total energy budget by virtue of its larger mass, as confirmed by the time-dependent bolometric luminosity shown in Fig.~\ref{Fig: Lbol_angle}.

\subsubsection{Light curves and comparison with GRB~230307A}

\begin{figure*}
	\centering
	\hspace{-8mm}
	\includegraphics[width=1\textwidth]{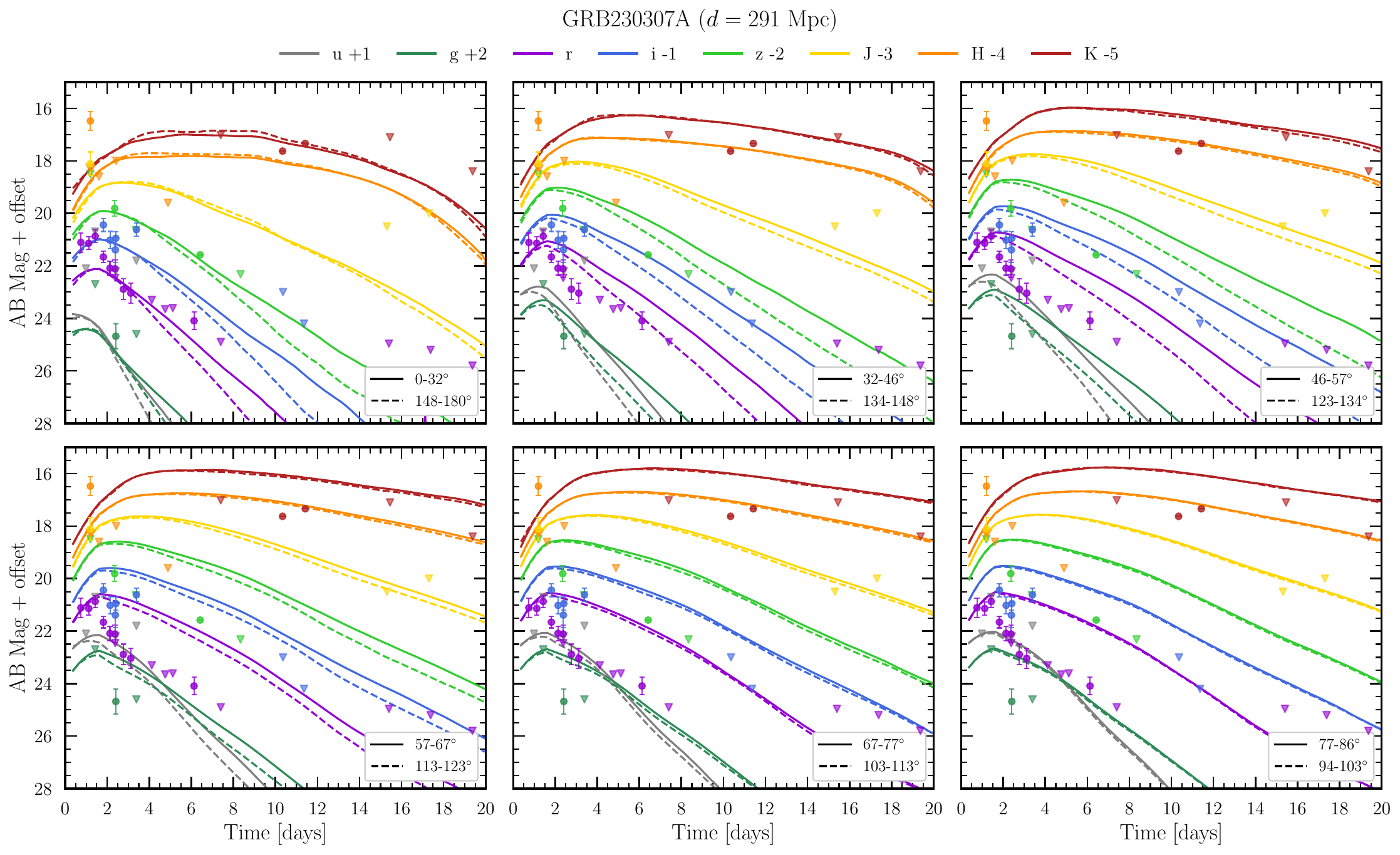}
	\caption{Synthetic broadband light curves in \texttt{ugrizJHK} bands for six pairs of viewing-angle bins, compared with the photometric observations of AT~2023vfi/GRB~230307A at a luminosity distance of $291$~Mpc, with data compiled from~\citet{Rastinejad:2024zuk}. Solid lines correspond to the northern hemisphere and dashed lines to the southern counterpart (e.g., $0^\circ$--$32^\circ$ and $148^\circ$--$180^\circ$). Data points show detections (circles with error bars) and upper limits (downward triangles). The mild north--south asymmetry visible in the polar bins reflects the independent evolution of the two hemispheres in our simulation and diminishes toward the equator. The best agreement with the data is achieved for the two most polar bins ($\theta_{\rm obs} \lesssim 30^\circ$), consistent with the near-polar viewing angle required by the GRB detection.}
	\label{Fig: lightcurves_GRB230307A}
\end{figure*}

Fig.~\ref{Fig: lightcurves_GRB230307A} presents the synthetic broadband light curves computed by \texttt{Sedona} in color bands for six pairs of viewing-angle bins, compared with the photometric observations of the kilonova AT~2023vfi associated with GRB~230307A at a luminosity distance of $291$~Mpc, with data compiled from~\citet{Rastinejad:2024zuk}. The color light curves are computed by convolving the spectral time series with the LSST \texttt{ugrizy} and JWST NIRCam \texttt{JHK} filter transmission curves.
Each panel shows the northern hemisphere (solid lines) and its southern counterpart (dashed lines) for a given angular bin: for example, $\theta_{\rm obs} = 0^\circ$--$32^\circ$ (solid) and $148^\circ$--$180^\circ$ (dashed).
Because the simulation does not impose north--south symmetry, these two hemispheres evolve independently.
A mild asymmetry is visible in the most polar bins, where one hemisphere produces slightly faster and less massive ejecta, leading to modestly brighter and faster-evolving light curves; the difference diminishes progressively toward the equator, where the two hemispheres are nearly indistinguishable.

\begin{figure}
	\centering
	\includegraphics[width=\columnwidth]{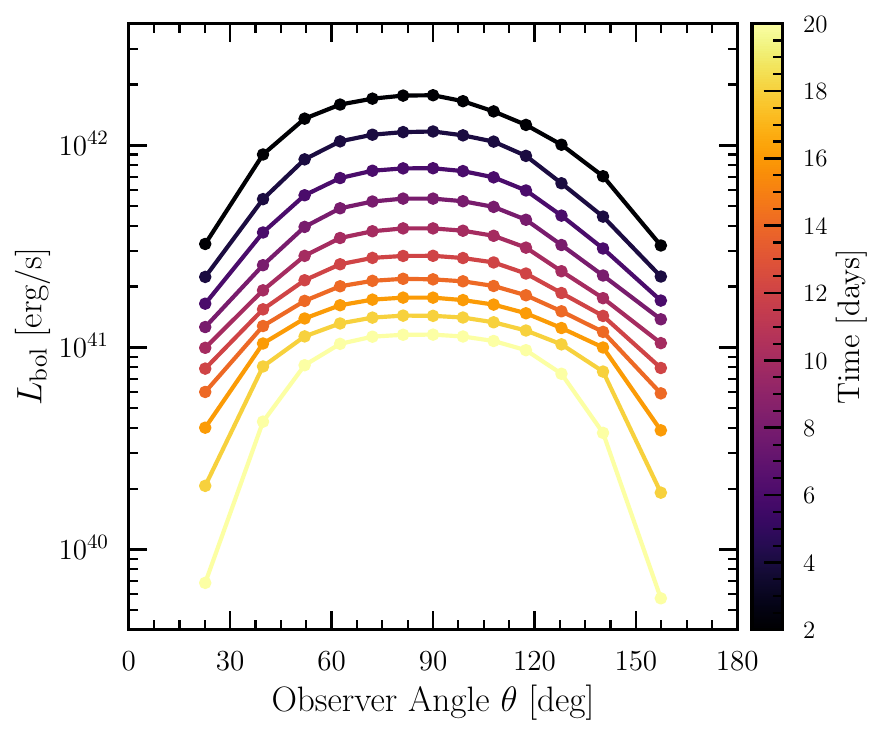}
	\caption{Bolometric luminosity as a function of observer angle $\theta_{\rm obs}$ at different epochs (color bar). The luminosity peaks near the equator at all times, reflecting the larger ejecta mass at mid-latitudes. The polar regions fade more rapidly due to the lower mass and faster expansion, which causes the ejecta to become optically thin on shorter timescales.}
	\label{Fig: Lbol_angle}
\end{figure}

The light curves exhibit a pronounced angular dependence across all bands (Fig.~\ref{Fig: lightcurves_GRB230307A}), as a direct consequence of the ejecta structure found in the \texttt{Gmunu} simulation (Sec.~\ref{sec:gmunu results}).
This is shown explicitly by the bolometric luminosity in Fig.~\ref{Fig: Lbol_angle}: at all times, $L_{\rm bol}$ peaks near the equator and drops steeply toward the poles, consistent with the concentration of ejecta mass at mid-latitudes (cf.\ right panel of Fig.~\ref{Fig: input_sedona_profile}).
The polar light curves also fade significantly faster, as the lower ejecta mass combined with the high expansion velocities causes the material to become optically thin on shorter timescales. This viewing-angle dependence is consistent with the geometric interpretation of~\citet{2020ApJ...897..150D}, who showed that for globally aspherical ejecta the observed luminosity scales primarily with the projected photospheric surface area along the line of sight; our equatorially concentrated ejecta can be roughly approximated as an oblate ellipsoid, for which equatorial observers see a larger projected area and hence brighter emission.

The best agreement with the AT~2023vfi photometry is achieved for the two most polar viewing-angle bins ($\theta_{\rm obs} \lesssim 32^\circ$ and $\theta_{\rm obs} \gtrsim 148^\circ$), where the model reproduces the observed fluxes without any parameter tuning---the ejecta mass, velocity structure, composition, and heating rates are all self-consistently determined by the simulation pipeline, with the viewing angle being the only free parameter.
Notably, the near-polar viewing angle required to match the kilonova photometry is the same geometry needed for the detection of prompt $\gamma$-ray emission, providing a non-trivial self-consistency check. 

\begin{figure}
	\centering
	\hspace{-8mm}
	\includegraphics[width=0.5\textwidth]{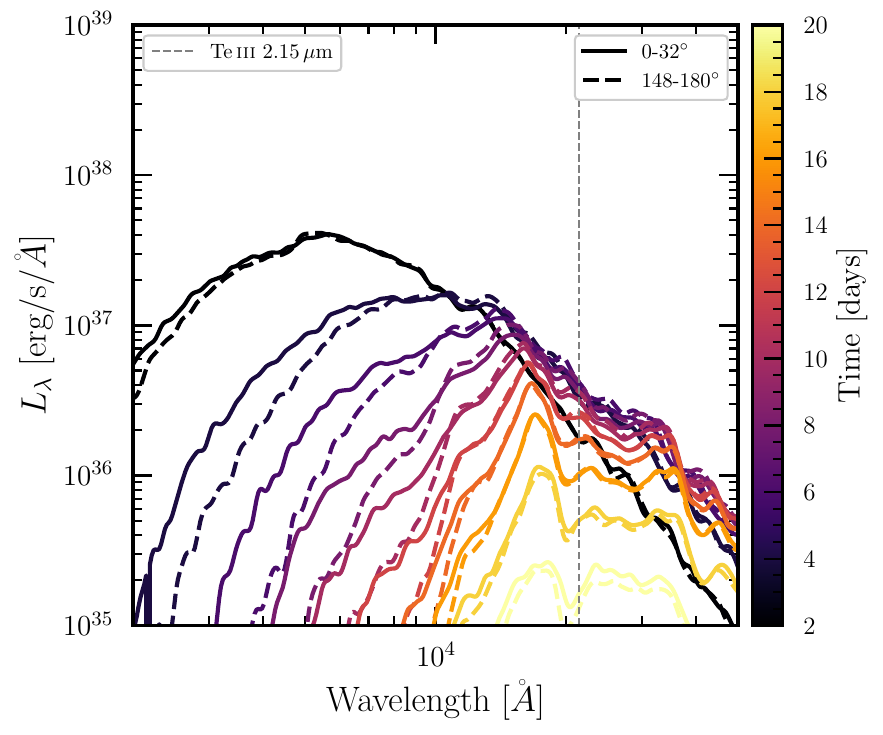}
	\caption{Synthetic spectra for the two polar viewing-angle bins: $0^\circ$--$32^\circ$ (solid) and $148^\circ$--$180^\circ$ (dashed), at epochs from ${\sim}\,2$ to $20$~days (color bar). The two hemispheres produce comparable spectra at all times. The vertical dashed line marks the rest wavelength of the [Te~{\sc iii}] $2.15~\mu$m transition; an emission feature emerges in this wavelength region at late times but is not centered on the [Te~{\sc iii}] rest wavelength (see text).}
	\label{Fig:spectra_polar_north_south}
\end{figure}

Fig.~\ref{Fig:spectra_polar_north_south} shows the synthetic spectra for the two polar viewing-angle bins ($0^\circ$--$32^\circ$, solid; $148^\circ$--$180^\circ$, dashed) at epochs ranging from ${\sim}\,2$ to $20$~days.
The spectra from the two hemispheres closely resemble each other at all times, confirming the near north--south symmetry already seen in the broadband light curves.
The spectral energy distribution shifts progressively to redder wavelengths as the photosphere recedes into the heavier inner layers, with broad absorption and emission features characteristic of lanthanide-rich ejecta.

Our spectra display broad emission structure in the ${\sim}1.5$--$2.5~\mu$m region,
in the vicinity of the spectral feature observed by JWST in AT~2023vfi at $+29$~days~\citep{2024Natur.626..737L,2024Natur.626..742Y} (see also~\citet{Gillanders:2024hcb}). \citet{2024Natur.626..737L} attributed the observed feature to the forbidden [Te~{\sc iii}] $\lambda 21050$~\AA\ transition, identifying it as a heavy neutron-capture element in a compact object merger. In our synthetic spectra, the emission in this wavelength range likely arises from a blend of many lanthanide and actinide lines rather than from a single identifiable transition, and its peak does not coincide with the [Te~{\sc iii}] rest wavelength (vertical dashed line in Fig.~\ref{Fig:spectra_polar_north_south}). As emphasized by~\citet{Tanaka:2019iqp}, the spectral features produced by the \texttt{HULLAC} atomic data set should not be used to identify individual elements in the spectra of real kilonovae: the theoretical transition wavelengths have not been calibrated to experimental values, and the ensemble of transitions represents the general statistical properties of the opacity rather than precise line positions. Therefore, while the presence of broad emission near $2~\mu$m is consistent with the heavy $r$-process composition of our ejecta, we refrain from drawing conclusions about specific elemental identifications from our synthetic spectra. 

While the canonical interpretation of GRB~230307A invokes a neutron star merger progenitor~\citep{2024Natur.626..737L}, the remarkable consistency between our self-consistent AIC model and the observations---without any parameter tuning---suggests that AIC deserves serious consideration as an alternative progenitor channel for this event.
Future observations---particularly of the host galaxy environment, the presence or absence of a surviving compact remnant, and the detailed spectral evolution---may help distinguish between these two scenarios. Most decisively, a coincident gravitational wave detection with the inspiral chirp signature characteristic of a compact binary merger would unambiguously rule out the WD collapse channel for future similar events.

\subsubsection{Radio transparency of the ejecta}
\label{sec:radio}

The formation of a long-lived magnetar remnant following AIC opens the possibility that the central engine could produce coherent radio emission, such as a fast radio burst (FRB), at some point during or after the explosion~\citep{2019ApJ...886..110M,Kremer:2023utr}.
Whether such a signal can escape depends on the optical depth of the overlying ejecta to radio frequencies. We compute here the free-free optical depth and dispersion measure (DM$_{\rm{ej}}$) through the AIC ejecta as a function of time and viewing angle.

For a radial ray at polar angle $\theta$, the free-free optical depth in the Rayleigh--Jeans limit ($h\nu \ll k_{\rm B}T$) is~\citep{1979rpa..book.....R}
\begin{equation}
    \tau_{\rm ff}(\nu, \theta, t) = \int_0^\infty \alpha_{\rm ff}\, {\rm d}s \,,
    \label{eq:tau_ff}
\end{equation}
where the absorption coefficient is
\begin{equation}
    \alpha_{\rm ff} = 0.018 \, g_{\rm ff} \, \bar{Z}^{\,2} \, n_e \, n_{\rm ion} \, T^{-3/2} \, \nu^{-2} \quad [\mathrm{cm}^{-1}]\,,
    \label{eq:alpha_ff}
\end{equation}
with $n_e$ and $n_{\rm ion}$ the electron and ion number densities, $T$ the gas temperature, $\nu$ the radio frequency, $\bar{Z} = n_e / n_{\rm ion}$ the mean ionic charge, and $g_{\rm ff} \approx 1.2$ the free-free Gaunt factor. All quantities are in CGS units.
The dispersion measure is
\begin{equation}
    {\rm DM_{ej}}(\theta, t) = \int_0^\infty n_e\, {\rm d}s  \,.
    \label{eq:DM}
\end{equation}

We evaluate these integrals using the self-consistent thermodynamic state computed by \texttt{Sedona}.
At each output time, the grid provides the gas temperature $T$, free electron density $n_e$, and mass density $\rho$, all evolved under homologous expansion and radiative equilibrium with the local radiation field.
The ion number density is obtained from $n_{\rm ion} = \rho / (\bar{A}\, m_{\rm H})$, where $\bar{A} = 1/\sum_i (X_i/A_i)$ is the number-averaged mean atomic mass computed from the full 113-species composition.
The ejecta become transparent to radio emission at frequency $\nu$ along direction $\theta$ when $\tau_{\rm ff}$ drops below unity.

\begin{figure}
	\centering
	\includegraphics[width=0.9\columnwidth]{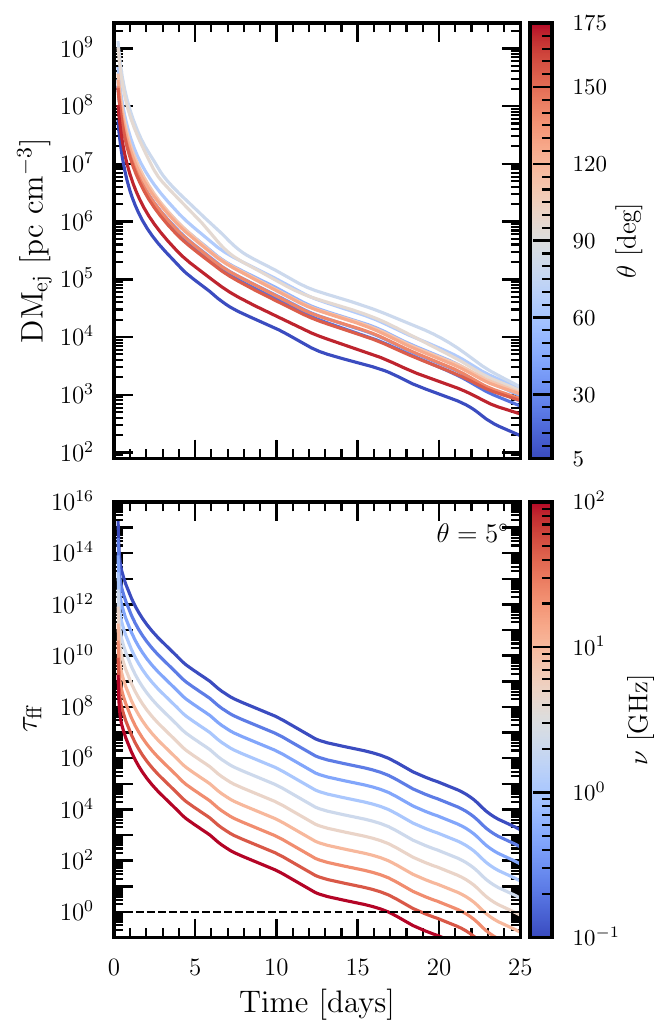}
	\caption{Radio transparency of the AIC ejecta. \textit{Upper panel:} Dispersion measure through the ejecta as a function of time for viewing angles from $\theta = 5^\circ$ (pole, blue) to $175^\circ$ (red). \textit{Lower panel:} Free-free optical depth along the polar direction ($\theta = 5^\circ$) for radio frequencies $0.1$ to $100$~GHz (color bar). The dashed line marks $\tau_{\rm ff} = 1$.}
	\label{Fig:radio}
\end{figure}

Fig.~\ref{Fig:radio} summarizes the results.
The upper panel shows the ejecta DM as a function of time for a representative set of viewing angles.
There is a clear angular dependence: the polar directions have the lowest column density, while equatorial lines of sight  carry the highest DM, reflecting the concentration of ejecta mass near the orbital plane.
By $t = 25$~days, 
the DM ranges from ${\sim}\,10^2~{\rm pc\,cm^{-3}}$ along the poles to ${\sim}\,10^3~{\rm pc\,cm^{-3}}$ near the equator.

The lower panel of Fig.~\ref{Fig:radio} shows $\tau_{\rm ff}$ along the polar direction ($\theta = 5^\circ$) for radio frequencies spanning $0.1$--$100$~GHz.
The highest frequencies become transparent first: at $\nu \sim 100$~GHz, $\tau_{\rm ff}$ drops below unity at $t \approx 20$~days.
All frequencies $\nu \gtrsim 5$~GHz become transparent along the pole within the simulation window.  Extrapolating the power-law decline beyond the simulation, we estimate that all viewing angles become transparent at ${\sim}\,$GHz frequencies within $t_{\rm trans} \lesssim 40$~days, with the polar directions clearing first.

We caution that these estimates rely on the assumption of local thermodynamic equilibrium (LTE) for the gas temperature and the Saha equation for the ionization state, as computed by \texttt{Sedona}.
At late times ($t \gtrsim 10$--$15$~days), the ejecta density drops sufficiently that LTE may no longer hold: the radiation field decouples from the gas, and non-LTE effects can alter both the temperature and the ionization balance.
In particular, the true free electron density may differ from the Saha prediction, which would directly affect $\tau_{\rm ff}$ and DM$_{\rm ej}$.
The transparency times and DM values quoted above should therefore be regarded as order-of-magnitude estimates (see \citealt{Brethauer:2025plw}, who showed that non-thermal ionization by radioactive products keeps the ejecta ionized well beyond the LTE prediction, which would increase both $\tau_{\rm ff}$ and DM$_{\rm ej}$).

With this caveat, the results indicate that any radio emission from the AIC remnant would be absorbed by the ejecta at early times, with the polar direction providing the first transparent window.
The $\sim$month-timescale radio transparency has important implications for multi-messenger follow-up strategies: radio observations of AIC candidates should target $t \gtrsim 30$~days at ${\sim}$GHz frequencies, with earlier detection possible at $\nu \gtrsim 10$~GHz along polar lines of sight.

\section{DISCUSSION}
\label{sec:discussion}

The nucleosynthesis yields presented in this work differ qualitatively from all previous predictions for AIC ejecta.
To place our results in context, we first review the physical reason why earlier studies found proton-rich outflows, then summarize existing nucleosynthesis predictions and hydrodynamic studies.

During the collapse, electron captures on protons drive the forming accretion disk to very low electron fractions ($Y_e \sim 0.1$).
However, sustained electron-neutrino irradiation of matter that remains bound near the proto-neutron star and its viscously spreading disk efficiently converts neutrons back to protons, raising $Y_e$ toward ${\sim}0.5$ by the time weak interactions freeze out~\citep{2009MNRAS.396.1659M}.
This re-leptonization is the fundamental reason why most previous AIC studies predicted proton-rich, $^{56}$Ni-dominated ejecta rather than $r$-process material.

Among the few studies that computed actual nucleosynthesis yields, \citet{2009MNRAS.396.1659M} modeled the disk wind evolution analytically and found that it synthesizes up to a few $\times 10^{-2}~M_\odot$ of $^{56}$Ni, with negligible $r$-process production.
\citet{Darbha:2010} built on this composition to compute radiative transfer, predicting fast, faint optical transients dominated by nickel features---a picture entirely different from the lanthanide-rich kilonova we find here.
\citet{Yip:2024akb} performed the first dedicated nucleosynthesis calculation from AIC hydrodynamic simulations, post-processing 1D unmagnetized models with a ${\sim}1200$-isotope network.
Their ejecta, with $Y_e \gtrsim 0.46$, produced only first neutron-capture peak elements (Sr, Y, Zr) and no heavy $r$-process---consistent with the expectation that neutrino irradiation has fully re-leptonized the outflow in the absence of magnetic fields.

Separately, several hydrodynamic studies have characterized the ejecta conditions ($Y_e$, entropy, velocity) without evolving a nuclear network.
The 2D MHD simulations of~\citet{2007ApJ...669..585D} showed that magnetic fields significantly enhance both the explosion energy and the ejected mass compared to unmagnetized models~\citep{2006ApJ...644.1063D}---in qualitative agreement with our results---and identified a bimodal $Y_e$ distribution with a neutron-rich component suggestive of $r$-process nucleosynthesis, though no nuclear network was evolved to quantify the resulting yields.
\citet{Batziou:2024ory} ran long-duration ($>5$~s) 2D simulations with detailed neutrino transport but without magnetic fields, finding $\langle Y_e \rangle \sim 0.43$--$0.50$.
The 3D simulations of~\citet{Kuroda:2025iyj}, which included rapid rotation and multi-energy neutrino transport with a weaker magnetic field ($B_0 = 10^{11}$~G), found that rotation broadens the $Y_e$ distribution toward lower values ($Y_e \sim 0.2$--$0.4$ for their most rapidly rotating models), but the ejecta remain dominated by $Y_e \gtrsim 0.3$ material. This is consistent with the trend found in CH2024, where only weakly magnetized models ($B_{\rm pol} \leq 10^{10}$~G) produce neutron-poor ejecta ($\langle Y_e \rangle \approx 0.5$), the $B_{\rm pol} = 10^{11}$~G model shows a peak in the ejecta mass distribution shifting toward $Y_e \sim 0.3$ but still insufficient for heavy $r$-process, and only the $B_{\rm pol} = 10^{12}$~G model ejects a large mass of material with $Y_e \leq 0.25$.
The 3D GRMHD simulation of~\citet{Combi:2025yvs} demonstrated that magnetic field amplification via a magnetorotational dynamo can drive ejecta with $Y_e \sim 0.25$---comparable to our findings---confirming from a fully 3D perspective that magnetized AIC can produce neutron-rich outflows, though they did not evolve a nuclear network either.

The present work bridges this gap by coupling, for the first time, an in-situ nuclear network to the radiation hydrodynamics of ejecta from a strongly magnetized, multidimensional GR$\nu$MHD simulation of AIC, and following the resulting composition through to synthetic observables.


\section{CONCLUSIONS}
\label{sec:conclusions}

We have presented the first end-to-end calculation connecting the dynamical evolution of an accretion-induced collapse to observable electromagnetic signatures, combining 2D GR$\nu$MHD simulations (\texttt{Gmunu}), radiation hydrodynamics with in-situ nuclear reaction networks (\texttt{kNECnn}: \texttt{SNEC}+\texttt{SkyNet}), and Monte Carlo radiative transfer (\texttt{Sedona}) with spatially resolved, composition-dependent heating rates. Our main findings are as follows:
\begin{enumerate}
    \item \textit{Robust heavy $r$-process nucleosynthesis.} The strongly magnetized AIC ejecta undergo strong
    $r$-process nucleosynthesis, producing elements up to and beyond the third $r$-process peak ($A \gtrsim 195$). The mass-weighted abundance pattern closely reproduces the solar $r$-process residuals across the second peak, rare-earth region, third peak, and actinides, while significantly underproducing first-peak elements, see Fig.~\ref{Fig: abundaces}. The four most abundant elements by mass---Xe, He, Pt, and Te---account for more than $50\%$ of the total ejecta.

    \item \textit{The essential role of magnetic fields.} Strong magnetic fields fundamentally alter the nucleosynthesis outcome of AIC by ejecting material on dynamical timescales, before neutrino irradiation can re-leptonize the outflow. This preserves the low $Y_e$ ($\sim 0.1$--$0.25$) produced during the collapse, enabling heavy element production that is qualitatively different from the $^{56}$Ni-dominated composition predicted by all previous unmagnetized models.

    \item \textit{Lanthanide-rich kilonova emission.} The lanthanide mass fraction of $X_{\rm lan} \approx 8\%$ produces a kilonova dominated by near-infrared emission that peaks at $\sim 4$~days, in contrast to the fast blue transients predicted by earlier unmagnetized AIC models.

    \item \textit{Spatially varying nuclear heating.} The thermalized heating rates can vary by up to an order of magnitude across the ejecta, driven by the spacial dependence of the nucleosynthesis. This highlights the importance of using composition-resolved heating in radiative transfer calculations, rather than uniform analytic prescriptions.

    \item \textit{Striking agreement with GRB~230307A.} Our synthetic light curves are in striking agreement with the photometric observations of the kilonova AT~2023vfi for polar viewing angles ($\theta_{\rm obs} \lesssim 30^\circ$), without any parameter tuning. The self-consistent requirement of a near-polar viewing angle---necessary both for the GRB detection and for the kilonova light curve match---provides
    strong support for AIC as a viable progenitor of this event. This identification is further supported by the long duration of GRB~230307A ($T_{90} \approx 33$~s)
    , which is more compatible with the spindown timescale of our AIC (${\mathcal{O} \sim10\rm{s}}$) than by a typical binary neutron star merger, and by the relatively large ejecta mass, which lies at the upper end of expectations for neutron star mergers.

\end{enumerate}
These results demonstrate that magnetized AICs can produce observable kilonova signals that are quantitatively comparable to those from neutron star mergers, establishing AIC as a potentially important---and observationally accessible---channel for heavy element production in the Universe.

Important directions for future work include extending the GR$\nu$MHD simulations to full 3D, where non-axisymmetric instabilities such as spiral modes and the MRI can develop self-consistently and may substantially affect the ejecta mass and geometry; exploring the dependence on the progenitor magnetic field strength, topology, and rotation rate; improving the initial WD models with self-consistent stellar evolution profiles; and developing non-LTE radiative transfer with experimentally calibrated atomic data to enable robust spectral diagnostics that can distinguish AIC-powered kilonovae from those produced by compact binary mergers.

\section*{Acknowledgements}
TP acknowledges support from NSF Grant PHY-2020275 (Network for Neutrinos, Nuclear Astrophysics, and Symmetries (N3AS)).
DR acknowledges support from the Sloan Foundation, from the Department of Energy, Office of Science, Division of Nuclear Physics under Awards Number DOE DE-SC0021177 an DE-SC0024388, from the National Science Foundation under Grants No. PHY-2020275, PHY-2407681, and PHY-2512802.
DK is supported in part by the U.S. Department of Energy, Office of Science, Office of Nuclear Physics, DE-AC02-05CH11231, DE-SC0004658, and DE-SC0024388, and by a grant from the Simons Foundation (622817DK).
FM acknowledges support from the Deutsche
Forschungsgemeinschaft (DFG) under Grant No. 406116891
within the Research Training Group RTG 2522/1. 
SB acknowledges support by the EU Horizon under ERC Consolidator Grant InspiReM-101043372 and support from the Deutsche Forschungsgemeinschaft, DFG, project MEMI number BE 6301/2-1.

The numerical simulations were performed on Perlmutter using NERSC award ERCAP0031370. This research used resources of the National Energy Research Scientific Computing Center, a DOE Office of Science User Facility supported by the Office of Science of the U.S.~Department of Energy under Contract No.~DE-AC02-05CH11231.

\section*{Data availability}
The nucleosynthesis yields, spectra, and broadband lightcurves from this simulation are publicly available at~\citet{pitik_2026_18758511}.


\bibliography{refs}


\appendix

\section{Nuclear network initialization}
\label{app:initialization}

\begin{figure*}
	\centering
	\includegraphics[width=0.85\textwidth]{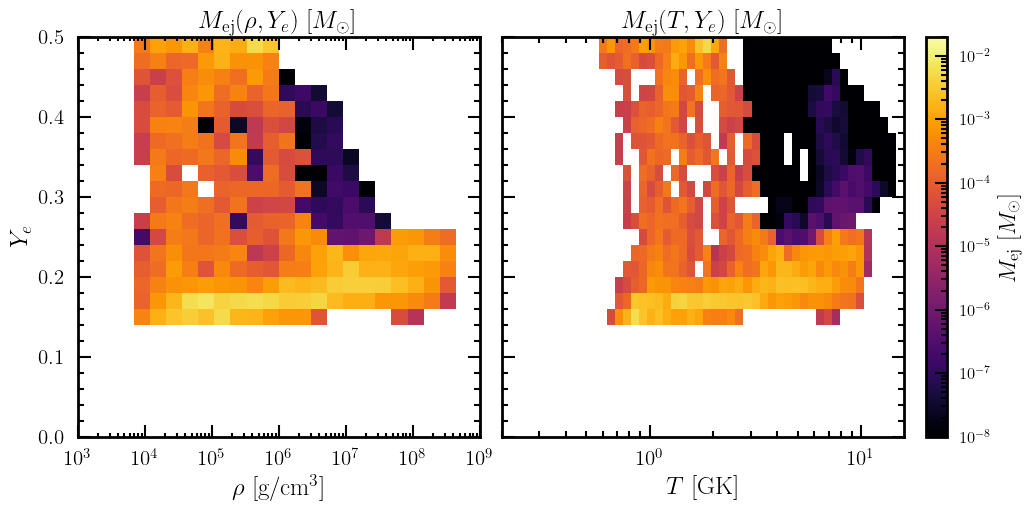}
	\caption{Joint distribution of unbound ejecta mass
			at $t_{\rm pb}=1.1$~s as a function of electron fraction $Y_e$
			versus rest-mass density $\rho$ (left) and $Y_e$ versus temperature
			$T$ (right). Most of the ejecta has expanded to
			$\rho\lesssim 10^7$~g~cm$^{-3}$ and cooled below
			$T_{\rm NSE}=8$~GK by extraction, so the backtracking-NSE
			initialization is applied to the bulk of the ejecta.}
	\label{Fig: ye_rho_T_hist}
\end{figure*}

\texttt{Gmunu} is coupled to a finite-temperature,
NSE-based equation of state (LS220 in our case), and evolves the
electron fraction $Y_e$ through the free-nucleon weak interactions
together with the two-moment neutrino transport. Individual nuclei are
not tracked in-simulation, so the $Y_e$ at extraction corresponds, to
a good approximation, to the weak-freeze-out value of each fluid
element. The post-NSE phase of the nuclear evolution---the assembly of
nuclei once NSE breaks down at $T\lesssim 5$--$8$~GK, their
$\beta^\pm$-decays along the $r$-process path, and the resulting drift
of $Y_e$ towards higher values---is recovered in post-processing along
the extracted trajectories \citep{2015MNRAS.448..541J,Radice:2018pdn,Magistrelli:2025xja}.

In our pipeline this handoff is realized through the
backtracking-NSE procedure introduced in
Section~\ref{sec:nucleosynthesis}: shells already below
$T_{\rm NSE}=8$~GK at extraction have their thermodynamic trajectory
extrapolated backwards under the assumption of homologous expansion to
the last time they crossed $T_{\rm NSE}$, NSE is imposed there using
the extraction $Y_e$ and entropy, and the network is then advanced
forward through the (extrapolated) trajectory before resuming the
actual coasting evolution. For the bulk of the ejecta, which crosses
$T_{\rm NSE}$ well outside the neutrinosphere where the weak rates lie
below the local expansion rate, $Y_e$ is not expected to evolve
significantly between the backtracked NSE point and extraction;
similarly, the matter is in adiabatic expansion so the entropy is
preserved.

Figure~\ref{Fig: ye_rho_T_hist} shows the joint mass
distribution of the ejecta in the $(\rho, Y_e)$ and $(T, Y_e)$ planes
at the extraction time. The bulk of the material has expanded to
$\rho\lesssim 10^7$~g~cm$^{-3}$ and cooled well below $8$~GK by the
end of the simulation, so the backtracking step is applied to most of
the ejecta, and a non-negligible fraction has pre-evolution times
${\sim} 0.5$--$1$~s---longer than the typical extraction times of BNSM
simulations and outside the regime in which the procedure has been
benchmarked in detail~\citep{Magistrelli:2025xja}. For these shells
the network effectively follows the homologous extrapolation rather
than the actual GR$\nu$MHD trajectory, increasing the per-shell
initialization uncertainty. Comparisons of post-processing pipelines
based on Lagrangian tracer particles~\citep{Radice:2018pdn} and on
different NSE-initialization prescriptions~\citep{Magistrelli:2025xja}
typically find the integrated abundance pattern to be reproduced to
within a factor of ${\sim} 2$ across the full mass range, with the
largest variations confined to specific mass regions and per-shell
$Y_e$ differences at the ${\sim} 25$--$40\%$ level. The integrated $Y_e$
distribution at extraction (Fig.~\ref{Fig: 1D his}) is a direct
GR$\nu$MHD output and does not depend on the post-processing pipeline;
the detailed abundance pattern, including the relative weights of the
second and third $r$-process peaks, the lanthanide mass fraction, and
the actinide tail, is more sensitive to the procedure.

Coupling a full nuclear reaction network in-situ to
the GR$\nu$MHD evolution would remove these systematics altogether,
but is computationally prohibitive in multi-dimensional simulations.
The recent RHINE scheme of~\citet{Just:2025hyy} achieves an equivalent
result by emulating the relevant nuclear source terms with neural
networks trained on full-network calculations, and we plan to adopt
it in 3D follow-up AIC simulations. Lagrangian tracer particles and
approximate in-situ networks provide complementary intermediate steps
in the same direction.


\bsp	
\label{lastpage}
\end{document}